\documentclass[11pt]{article}
\usepackage{jcappub,natbib,float,caption,comment,bm}
\title{Ray-tracing log-normal simulation for weak gravitational lensing: application
to the cross-correlation with galaxies}
\author[a]{Ryu Makiya,}
\author[b]{Issha Kayo,}
\author[c,a]{Eiichiro Komatsu}
\affiliation[a]{Kavli Institute for the Physics and Mathematics of the
Universe, Todai Institutes for Advanced Study, the University of Tokyo,
Kashiwa, Japan 277-8583 (Kavli IPMU, WPI)}
\affiliation[b]{Department of Liberal Arts, Tokyo University of Technology,
5-23-22 Nishikamata, Ota-ku, Tokyo, Japan}
\affiliation[c]{Max-Planck-Institut f\"ur Astrophysik, Karl-Schwarzschild-Str.
1, 85741 Garching, Germany}
\emailAdd{ryu.makiya@ipmu.jp}
\emailAdd{kayouissha@stf.teu.ac.jp}
\emailAdd{komatsu@mpa-garching.mpg.de}
\abstract{
We present an algorithm to self-consistently generate mock weak gravitational lensing convergence fields and galaxy distributions in redshift space. We generate three-dimensional cosmic density fields that follow a log-normal distribution, and ray-trace them to produce convergence maps. As we generate the galaxy distribution from the same density fields in a manner consistent with ray-tracing, the galaxy-convergence cross-power spectrum measured from the mock agrees with the theoretical expectation with high precision. 
We use this simulation to forecast the quality of galaxy-shear cross-correlation measurements from the Subaru 
Hyper Suprime-Cam (HSC) and Prime Focus Spectrograph (PFS) surveys.
We find that the nominal HSC and PFS surveys would detect the cross power spectra with signal-to-noise ratios of 20  and 5 at the lowest ($z = 0.7$) and highest ($z = 2.2$) redshift bins, respectively.
}
\begin{document}
\maketitle
\flushbottom

\section{Introduction}
The large-scale structure (LSS) of the universe is a powerful tool for cosmology \cite{peebles:1980}. 
It has been intensively studied using various probes such as galaxy clustering and weak gravitational lensing shear fields. 
See refs.~\cite{alam/etal:2017,alam/etal:2020,troxel/etal:2018,hikage/etal:2019,hildebrandt/etal:2020,heymans/etal:2020} for recent measurements.

The galaxy clustering in redshift space, mainly measured from spectroscopic galaxy samples, offers a probe of the expansion history of the universe as well as the growth rate of the structure through the baryon acoustic oscillations \cite{eisenstein/etal:2005,cole/etal:2005} and the redshift space distortion (RSD) \cite{jackson:1972,sargent/turner:1977,kaiser:1987}.
A key ingredient in the analysis of the galaxy clustering is a galaxy bias (see \cite{desjacques/etal:2018} for a review), which relates the clustering amplitude of galaxies to the underlying dark matter density fields. 
The galaxy bias is usually treated as nuisance parameters, which limit the constraining power of the galaxy clustering on cosmological parameters.

The cosmological weak gravitational lensing effect is a magnification and coherent distortion of galaxy images induced by the intervening matter density field \cite{schneider/ehlers/falco:1992}.
Unlike the galaxy clustering, the weak lensing effect offers a measure of the total matter density field free from the galaxy bias, since it is purely gravitational. 
It also allows us to study the expansion history and the growth of matter density fields.
However, one of the disadvantages of the cosmological weak lensing effect is its low redshift resolution.
Since the amplitude of the lensing power spectrum is determined by the line-of-sight integral of the matter density field, it is not straightforward to perform the ``tomographic'' analysis of the lensing data alone.

The galaxy clustering and the weak lensing effect are complimentary, as their joint analysis can lift degeneracy between the galaxy bias and the cosmological parameters.
The cross-correlation of the spectroscopic galaxy samples and the weak lensing effect enables us to perform the redshift tomography of the gravitational lensing shear and convergence fields. 
Furthermore, the multi-probe analysis is useful for finding potential systematics in each of the LSS surveys, to obtain robust results.

There are several on-going and planned LSS surveys aiming to observe the galaxy clustering and the weak lensing effect with unprecedented precision \cite{alam/etal:2017,alam/etal:2020,troxel/etal:2018,hikage/etal:2019,hildebrandt/etal:2020,heymans/etal:2020}. 
In this paper we take the weak lensing survey from the Subaru Hyper-Sprime Cam (HSC) \cite{hikage/etal:2019} and the spectroscopic galaxy redshift survey from the Prime Focus Spectrograph (PFS) \cite{takada/etal:2014} as an example. 
The HSC provides precise maps of the cosmic shear field, while the PFS, a fiber-fed multi-object spectrograph, will perform a spectroscopic galaxy survey on top of the HSC photometric galaxy samples. 
A joint analysis of the HSC and PFS data will provide new insights for cosmology.
Other planned LSS survey projects include Dark Energy Spectroscopic Instrument (DESI) \cite{desi}, Legacy Survey of Space and Time (LSST) by Vera C. Rubin Observatory \cite{lsst}, Nancy Grace Roman Space Telescope \cite{wfirst} and Euclid \cite{euclid}.

To extract robust cosmological results from the LSS surveys, it is important to understand statistical and systematic uncertainties both in the model and observations. 
An end-to-end simulation is an essential tool for understanding these uncertainties. 
To this end, $N$-body simulations have often been used. 
For example, the authors of ref.~\cite{takahashi/etal:2017} generated full-sky cosmic shear and convergence maps by ray-tracing $N$-body simulations of dark matter halos, and the mock HSC shear catalog generated from it has been used in the HSC cosmology analysis \cite{hamana/etal:2020}.
However, as LSS surveys become increasingly larger, the required number of simulations  also increases, demanding more computational resources.

The so-called ``log-normal simulation'' offers a computationally less expensive alternative. 
While this kind of simulations do not capture physics completely correctly, they are useful for capturing basic summary statistics such as the two-point correlation function (which is provided as the input to the simulation) and the associated covariance matrix \cite{chiang/etal:2013,alonso/etal:2014,Xavier/etal:2016,Agrawal/etal:2017,hand/etal:2018,blot/etal:2019,lippich/etal:2019}, and testing real-world issues such as the effects of a survey window function \cite{chiang/etal:2013}, interlopers \cite{addison/etal:2019}, and fiber collisions \cite{sunayama/etal:2020} via end-to-end simulations.

In the log-normal method, matter density fields are generated to follow the desired clustering properties instead of solving the gravitational evolution from initial conditions, assuming that their probability density function (PDF) follows a log-normal distribution. 
This method is motivated by the observation that the one-point PDF of log-transformed density fields measured from $N$-body simulations is approximately a Gaussian \cite{coles/etal:1991,colombi:1994,kofman/etal:1994,bernardeau/kofman:1995,uhlemann/etal:2016,shin/etal:2017}, and the two-point PDF also follows approximately a multivariate log-normal distribution \cite{kayo/etal:2001}.
The lognormal method, by design, generates the density fluctuation field whose one-point and two-point statistics are consistent with the $N$-body simulations, although the direct field-to-field comparison is not possible because the lognormal simulation does not solve the evolution of density fluctuation.

There are several other `approximate methods' for the gravitational structure formation \cite[see][for review]{monaco/2016}.
Ref. \cite{lippich/etal:2019,blot/etal:2019,colavincenzo/etal:2019} performed a systematic comparison of the clustering measurements of those `approximate methods', including the log-normal method, with a reference $N$-body simulation.
They show that the real space two-point clustering matches with the $N$-body and other approximate methods, although the bispectrum and the redshift space clustering measurements of the lognormal method show deviations from the $N$-body simulation.
They also show that the computational cost of the lognormal method is still less expensive than others, such as the method based on the Lagrangian perturbation theory; thus it is still useful to develop the lognormal code.
Another advantage of the log-normal method is that the statistical properties of the output fields are predictable from the inputs by design. 
This property makes it easier to test the real-world issues and systematics related to the observations (a few examples were mentioned above) without worrying about numerical uncertainties and computational costs associated with physical simulations.

In this paper we present a new code, {\tt lognormal\_lens}, which generates simulated distributions of the weak lensing convergence field and galaxies in redshift space from the common three-dimensional log-normal matter density field in a self-consistent manner. 
This code is based on publicly available {\tt lognormal\_galaxies}  \cite{Agrawal/etal:2017} and ray-tracing RAYTRIX \cite{Hamana:2001vz} codes.
We use {\tt lognormal\_galaxies} to generate matter density, velocity, and galaxy density fields from the input power spectrum, and compute galaxy power spectra in redshift space.
We then construct a light cone from the matter density fields and ray-trace it using RAYTRIX to obtain a weak lensing convergence map.
The {\tt lognormal\_lens} also computes auto- and cross-power spectra of the weak lensing convergence and galaxy density fields, which enables us to measure the cross-covariance of galaxy power spectra in redshift space and cosmic shear two-point functions.
After validating the {\tt lognormal\_lens} code, we use it to generate mock data of the HSC and PFS and study how well they can measure the cross-correlation power spectra of weak lensing and galaxy density fields. 

Our new code is complementary to the existing log-normal codes such as FLASK for weak lensing \cite{Xavier/etal:2016} and CoLoRe \cite{alonso/etal:2014} for the 21-cm line intensity mapping.
FLASK, which is used for the analysis of the Dark Energy Survey cosmic shear data, generates all-sky maps of the galaxy density and cosmic shear and convergence fields.
Instead of ray-tracing the three-dimensional density fields, FLASK generates random realizations of the convergence field  on a two-dimensional sphere from input angular power spectra assuming that it follows a distorted log-normal PDF.
FLASK does not produce the velocity field; thus, the anisotropic galaxy power spectrum in redshift space is given as an input. CoLoRe generates mock observations of 21-cm line intensity mapping and RSD, but does not generate weak lensing shear fields.

The paper is organized as follows.
In Section \ref{sec:wl_basics} we  review basics of the weak gravitational lensing effect.
In Section \ref{sec:ln_sim} we describe our method for generating the weak lensing convergence field.
In Section \ref{sec:validation} we validate our algorithm by comparing the simulated auto- and cross-spectra with the theoretical predictions.
In Section \ref{sec:forecast} we present the forecast for the future Subaru PFS and HSC surveys.
We summarize and conclude in Section \ref{sec:summary}.
In Appendix \ref{sec:code_docs} we describe the modifications to RAYTRIX.

\section{Basics of the weak gravitational lensing effect}
\label{sec:wl_basics}
\subsection{Shear and convergence fields}
The weak lensing effect is described as transformation from the unlensed ($\vec{\theta}^{u}$) to lensed ($\vec{\theta}$) coordinates.
With the so-called Born approximation, this transformation is written as
\begin{equation}
    \vec{\theta}^{u} = \vec{\theta}-\vec{\nabla}\varphi,
\end{equation}
where $\vec{\nabla} = \partial/\partial \vec{\theta}$ is a two-dimensional gradient in the angular axis.
The lens potential $\varphi$ is defined as
\begin{equation}
    \label{eq:lens_potential}
    \varphi(\vec{\theta}) = 
    \frac{2}{c^2 \chi_s}
    \int^{\chi_s}_0 {\rm d}\chi\;
    \frac{\chi_s-\chi}{\chi}
    \Phi(\chi,\vec{\theta}),
\end{equation}
where $\chi_s$ is the comoving distance to the source and $\Phi$ is the gravitational potential.
The Jacobian matrix of the lensed to unlensed coordinate transformation, $A$,  called the magnification matrix, is commonly written as
\begin{equation}
    \label{eq:kappa_gamma_jacobian}
    A = 
    \begin{pmatrix}
    1-\kappa-\gamma_1 & -\gamma_2 \\
    -\gamma_2 & 1-\kappa+\gamma_1
    \end{pmatrix},
\end{equation}
where the convergence $\kappa$ and shear fields $\gamma_1$, $\gamma_2$ are defined by
\begin{equation}
    \label{eq:def_kappa_gamma}
    \kappa = \frac{1}{2}\vec{\nabla}^2\varphi,\;
    \gamma_1 = \frac{1}{2}\left(
    \frac{\partial^2 \varphi}{\partial \theta_1^2}
    -\frac{\partial^2 \varphi}{\partial \theta_2^2}
    \right),\;
    \gamma_2 = \frac{\partial^2 \varphi}{\partial \theta_1 \partial \theta_2}.
\end{equation}
From eq.(\ref{eq:lens_potential}) and (\ref{eq:def_kappa_gamma}), $\kappa$ is given by
\begin{equation}
    \label{eq:kappa_phi}
    \kappa(\vec{\theta}) = 
    \frac{1}{c^2 \chi_s}
    \int^{\chi_s}_0 {\rm d}\chi\;
    \frac{\chi_s-\chi}{\chi}
    \vec{\nabla}^2
    \Phi(\chi,\vec{\theta}).
\end{equation}
In eq.(\ref{eq:kappa_phi}) we can replace $\vec{\nabla}^2$ with the three-dimensional
Laplacian $\bigtriangleup$, 
\begin{equation}
    \bigtriangleup = \frac{1}{\chi^2}\frac{\partial}{\partial \chi} 
    \left( \chi^2 \frac{\partial}{\partial \chi} \right)+
    \frac{1}{\chi^2}\vec{\nabla}^2,
\end{equation}
assuming that the positive and negative contributions cancel when we integrate $\partial^2/\partial \chi^2$ along the line-of-sight \cite[e.g.,][]{Bartelmann/Schneider:2001,Kilbinger:2015}.
Combining with the Poisson equation 
\begin{equation}
\bigtriangleup \Phi = \frac{3H_0^2 \Omega_{m0}}{2c^2}
\frac{\delta}{a},     
\end{equation}
we obtain
\begin{equation}
    \label{eq:kappa_delta}
    \kappa(\vec{\theta}) =
    \frac{3H_0^2\Omega_{m}}{2c^2}
    \int^{\chi_s}_0 {\rm d}\chi\;
    \frac{\chi}{a(\chi)}
    \left(1-\frac{\chi}{\chi_s}\right)
    \delta(\chi,\vec{\theta}),
\end{equation}
where $a$ is the cosmological scale factor.
This equation shows that convergence $\kappa$ can be interpreted as the line-of-sight integration of the density fluctuation field $\delta$ with a lensing weight function 
\begin{equation}
    \label{eq:wk_single}
    W_{\kappa}(\chi) = 
    \frac{3H_0^2\Omega_{m}}{2c^2}
    \frac{\chi}{a(\chi)}
    \left(1-\frac{\chi}{\chi_s}\right).
\end{equation}
If we consider $\kappa$ for multiple source galaxies with a redshift distribution given by ${\rm d}n_{\rm g,s}/{\rm d}\chi$, the lensing weight function is replaced by
\begin{equation}
    \label{eq:wk_multi}
    W_{\kappa}(\chi) = 
    \frac{3H_0^2\Omega_{m}}{2c^2}
    \int_{\chi}^{\infty}
    {\rm d}\chi_s\;
    \frac{{\rm d}n_{\rm g,s}}{{\rm d}\chi}(\chi_s)
    \frac{\chi}{a(\chi)}
    \left(1-\frac{\chi}{\chi_s}\right),
\end{equation}
and the integration range of eq.(\ref{eq:kappa_delta}) is altered to $[0,\infty]$.
Here the source distribution function is normalized to unity, i.e., $\int_0^{\infty} {\rm d}\chi\; {\rm d}n_{\rm g,s}/{\rm d}\chi = 1$.

In Fourier space the convergence and shear fields are expressed as
\begin{equation}
    \tilde{\kappa}(\vec{\ell}) =
-\frac{|\vec{\ell}|^2}{2}\tilde{\varphi}(\vec{\ell}),\;
    \tilde{\gamma}_1(\vec{\ell}) =
-\frac{\ell_1^2-\ell_2^2}{2}\tilde{\varphi}(\vec{\ell}),\;
    \tilde{\gamma}_2(\vec{\ell}) = -\ell_1 \ell_2 \tilde{\varphi}(\vec{\ell}),
\end{equation}
where the tildes denote Fourier transformed quantities and $\vec{\ell} = (\ell_1,\ell_2)$ is the wave-number vector. 
The convergence and shear fields are related by
\begin{equation}
    \tilde{\gamma}_1(\vec{\ell}) =
\cos{(2\phi_\ell)}\tilde{\kappa}(\vec{\ell}),\;
    \tilde{\gamma}_2(\vec{\ell}) =
\sin{(2\phi_\ell)}\tilde{\kappa}(\vec{\ell}),
\end{equation}
where $\phi_{\ell}$ is defined as $(\cos{\phi_{\ell}},\sin{\phi_{\ell}}) = (\ell_1/\ell,\ell_2/\ell)$.
If we consider the following coordinate rotation,
\begin{equation}
    \begin{pmatrix}
    \tilde{\gamma}_{E} \\
    \tilde{\gamma}_{B}
    \end{pmatrix}
    =
    \begin{pmatrix}
    \cos{2\phi_{\ell}} & \sin{2\phi_{\ell}}\\
    -\sin{2\phi_{\ell}} & \cos{2\phi_{\ell}}
    \end{pmatrix}
    \begin{pmatrix}
    \tilde{\gamma}_1\\
    \tilde{\gamma}_2
    \end{pmatrix}
    ,
\end{equation}
we find
\begin{equation}
    \label{eq:shear_convergence}
    \tilde{\gamma}_E = \tilde{\kappa},\;\tilde{\gamma}_{B} = 0.
\end{equation}
These shear components $\tilde{\gamma}_E$ and $\tilde{\gamma}_B$ are called
the $E$- and $B$-mode, respectively.

\subsection{Angular power spectra of the convergence field}
The power spectrum (or the two-point correlation function) is the commonly-used summary statistics of the observed cosmic shear field.
In the following we focus on the convergence rather than shear, since the relationship between the convergence and matter density fields is simpler,and the shear can be obtained from the convergence by eq.(\ref{eq:shear_convergence}) if there is no ambiguity in the $E$- and $B$-mode separation.

The convergence $\kappa$ is expressed as the line-of-sight integration of the matter density fluctuation $\delta$ weighted by the lensing kernel $W_{\kappa}$, and thus the angular power spectrum of $\kappa$ is also written as the weighted line-of-sight integration of the matter power spectrum $P_{\rm mm}$.
With the Limber approximation \cite{loVerde/afshordi:2008}, the angular auto-power spectrum of convergence
field is written as
\begin{equation}
\label{eq:clkk}
 C^{\kappa \kappa}(\ell) = 
 \int_0^{\infty} d\chi\;W_\kappa^2(\chi) \chi^{-2}
  P_{\rm mm}((\ell+0.5)/\chi, z).
\end{equation}
Analogously, the cross power spectrum of the galaxy density and  convergence fields $C_{\ell} ^{\kappa g}$ and the angular auto power spectrum of galaxies  $C_{\ell}^{gg}$ are written as 
\begin{equation}
\label{eq:clgk}
 C^{\kappa g}(\ell) = 
 \int_0^{\infty} d\chi\;W_\kappa(\chi)W_{g}(\chi) \chi^{-2}
  P_{\rm gm}((\ell+0.5)/\chi, z),
\end{equation}
and 
\begin{equation}
\label{eq:clgg}
 C^{gg}(\ell) = 
 \int_0^{\infty} d\chi\;(\chi)W_{g}^2(\chi) \chi^{-2}
  P_{\rm gg}((\ell+0.5)/\chi, z),
\end{equation}
where $P_{\rm gm}$ and $P_{\rm gg}$ are the galaxy-matter cross-power spectrum and the galaxy auto power spectrum, respectively.
The galaxy kernel $W_g$ is the radial distribution
of lens galaxies normalized to unity, ${\rm d}n_{\rm g}/{\rm d}\chi$.

\section{Log-normal simulation}
\label{sec:ln_sim}
The {\tt lognormal\_lens} code is based on our lognormal code {\tt lognormal\_galaxies}\footnote{The code is also publicly available at http://wwwmpa.mpa-garching.mpg.de/\textasciitilde komatsu/codes.html.}, which generates a density field on regular grids from an input power spectrum, or, equivalently, two-point correlation function, assuming that the density field follows the log-normal PDF as follows.

The code first generates the log-transformed density fluctuation field $\ln [1+\delta(\bm{x})]$ which follows the Gaussian random field characterized by the two-point correlation function $\xi^{G}(r)$. 
The Gaussian correlation function $\xi^{G}(r)$ is related to the input correlation function $\xi(r)$ as \cite{coles/etal:1991}
\begin{equation}
    \xi^{G}(r) = \ln{[1+\xi(r)]}.
\end{equation}
Then the density fluctuation field $\delta(\bm{x})$ is obtained by exponentiating the log-transformed field.
Note that this treatment naturally satisfies the physical constraint on the density contrast, $\delta > -1$, which is not the case for the Gaussian random field with large variance such as of galaxies.
We caution that the lognormal method is not guaranteed to reproduce the higher-order correlations of density field.
Indeed, Ref.\cite{colavincenzo/etal:2019} shows that the {\tt lognormal\_galaxies} does not reproduce the halo bispectrum measured from the $N$-body simulation.

For the matter density field we use the linear matter spectrum as an input, while for the galaxy density field we use the linear galaxy power spectrum which is obtained by multiplying the linear bias squared to the matter power spectrum.
We do not implement any non-linear galaxy bias into our simulations.
The mock galaxies are then Poisson sampled from the underlying galaxy density field with the mean number given as inputs.
We use the same random seed for the Gaussian field of galaxies and matter, to ensure that the galaxy density field and matter density field are correlated.
The code also generates the peculiar velocity field of galaxies from the underlying matter density field by using the linear continuity equation.

We refer the readers to ref.~\cite{Agrawal/etal:2017} for further details of {\tt lognormal\_galaxies}.

\subsection{Weak lensing convergence field}
\label{sec:sim_lens}
In {\tt lognormal\_lens} we ray-trace a light-cone of the matter density field to obtain the weak lensing convergence and shear fields. Ray-tracing is performed by the public code RAYTRIX \cite{Hamana:2001vz}.
As is done by Ref.\cite{Xavier/etal:2016}, the weak lensing fields can be approximated by the distorted lognormal fields without constructing light cone. However one of the our purpose is to measure the cross correlation between convergence field and galaxies, and therefore we need to construct the light cone from the matter density fields which correlate with the galaxy density field at the same redshift.

\begin{figure}[]
  \begin{center}
    \includegraphics[width=1.0\textwidth]{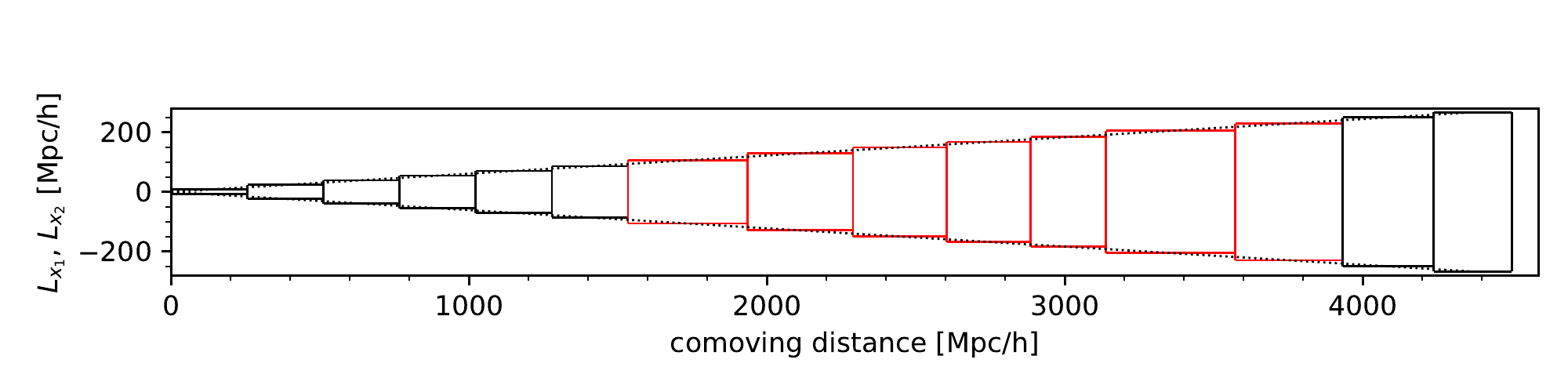}
    \caption{Configuration of the light cone with rectangular simulation boxes. The matter density field is generated in each box. 
The black dotted lines show the boundary of a $7\times7$ ${\rm deg}^2$ field of view.  We generate galaxies in boxes with the red boundaries, mocking spectroscopic samples.}
    \label{fig:lognormal_lens_geometry}
  \end{center}
\end{figure}

The code first generates the light cone from a series of three-dimensional matter density fields generated by the {\tt lognormal\_galaxies}, assuming that the density field does not evolve within the redshift width of the box and the different redshift boxes are not correlated. Since the lognormal method does not solve the evolution of density fields, we can not have a light cone with the smooth transition of density fields.
The light cone geometry of our fiducial simulation is shown in Figure \ref{fig:lognormal_lens_geometry}.

The side lengths of boxes perpendicular to the line of sight ($L_{x_1}$ and $L_{x_2}$ in Figure \ref{fig:lognormal_lens_geometry}) are chosen to match the opening angle of the simulation (shown in the black dotted lines) times comoving distances (the horizontal axis) to the box centers at various $z$.
Due to the limitation of RAYTRIX, we can only deal with a square map in angle space. 
In this setting the boundary of simulation boxes and the survey opening angle match; thus, we do not need to worry about the window effect due to the survey geometry.

The box length along the line of sight is chosen as follows.
We first create boxes in which spectroscopic galaxy samples are generated (the red boxes in Figure \ref{fig:lognormal_lens_geometry}). 
The positions and sizes of these boxes are determined by the assumed survey strategy.
In our fiducial simulation, spectroscopic galaxies are generated at $0.6 < z < 1.6$ with the redshift interval of $\Delta z = 0.2$ and at $1.6 < z < 2.4$ with $\Delta z = 0.4$, assuming the PFS survey parameters given in Table \ref{tb:pfs}.
The geometry of boxes located at lower and higher redshifts than the galaxy samples are chosen to have depths similar to those of the boxes of the galaxy samples. 
The maximum redshift $z_{\rm max}$ is chosen to match the assumed redshift distribution of source galaxies. 
Here we set $z_{\rm max} = 3.2$ assuming the HSC survey.

\begin{table}[]
\begin{center}
\begin{tabular}{ccccc}
Redshift & $V_{\rm survey}$ & $\bar{n}_g$ 	& bias & \\
		 & $[h^3\;{\rm Gpc}^{-3}]$ & $[10^{-4}\;h^3{\rm Mpc}^{-3}]$ 	& & \\ 
\hline
$0.6 < z < 0.8$ & 0.59 & 1.9 & 1.18 &  \\
$0.8 < z < 1.0$ & 0.79 & 6.0 & 1.26 &  \\
$1.0 < z < 1.2$ & 0.96 & 5.8 & 1.34 &  \\
$1.2 < z < 1.4$ & 1.09 & 7.8 & 1.42 &  \\
$1.4 < z < 1.6$ & 1.19 & 5.5 & 1.50 &  \\
$1.6 < z < 2.0$ & 2.58 & 3.1 & 1.62 &  \\
$2.0 < z < 2.4$ & 2.71 & 2.7 & 1.78 &  \\
\hline
\end{tabular}
\caption{The PFS cosmology survey parameters.}
\label{tb:pfs}
\end{center}
\end{table}	

We use the plane-parallel approximation to project the three-dimensional matter density field of the $i$-th box, $\delta_i$, onto the two-dimensional density field $\delta_i^{\rm proj}$ (hereafter we refer it to as ``mass sheet'') as
\begin{equation}
	\delta^{\rm proj}_{i}(x_1,x_2) =
	\frac1{N_{{\rm grid},y}}
	\sum_{y_i} \delta_{i}(x_1,x_2,y_i).
\end{equation}
We shall denote the comoving Cartesian coordinates of the grid center as $(x_1,x_2,y)$ with $y$ being the line-of-sight axis.
The summation is taken over all $y_i$, where $N_{{\rm grid},y}$ is the number of grids along the line of sight.
The number of grids along the $x_1$ and $x_2$ axes are set to be equal to that of the resultant angular map, which is given as the input parameter.
The number of grids along the $y$ axis is determined so that each grid cell becomes a cube.
Since the matter density field is generated to satisfy the periodic boundary condition, the mass sheet is also periodic.

Next, the code calculates the two-dimensional deflection potential of the $i$-th mass sheet, $\Psi^i$, via the Poisson equation as 
\begin{equation}
\label{eq:poisson}
	\nabla^2 \Psi^i(x_1,x_2) = \frac{3\Omega_m H_0^2}{c^2} \delta^{\rm
proj}_i(x_1,x_2).
\end{equation}
The position of light ray at the source plane, $\bm{\theta}^s$, which is at $\bm{\theta}^1$ on the image plane, is deflected as
\begin{equation}
    \bm{\theta}^s = \bm{\theta}^1-\sum_{i=1}^{n-1}
    \frac{(\chi_s-\chi_i)}{a(\chi_i)\chi_s}
    \nabla_{\perp}\Psi^{i},
    \label{eq:deflection}
\end{equation}
where $\nabla_{\perp}$ is $\partial/\partial x_1$ or $\partial/\partial x_2$, $\chi_i$ is the comoving distance to the $i$-th mass sheet, and $\chi_s$ is the comoving distance to the source plane. 
The source plane is located at the $n$-th mass sheet, i.e., $\chi_s = \chi_n$.

The light ray is propagated assuming the plane-parallel approximation; thus, the spatial position of the $i$-th plane, $\bm{x}^i$, is converted to the angular position $\bm{\theta}^i$ as  $\bm{\theta}^i = \bm{x}^i/\chi_i$.
The code first solves eq.(\ref{eq:poisson}) using the Fast Fourier Transform, and then evaluates the derivatives of $\Psi$ in eq.(\ref{eq:deflection}) using the finite difference method.

Along the light ray path, the Jacobian matrix of the lensed-to-unlensed coordinate transformation for sources at $\chi_s$, $A_s$, is calculated as 
\begin{equation}
    A_s = I-\sum_{i=1}^{n-1}
    \frac{\chi_i(\chi_s-\chi_i)}{a(\chi_i)\chi_s}
    U_i A_i,
\end{equation}
where $I$ is the identity matrix and 
\begin{equation}
    U_i = 
    \begin{pmatrix}
    \displaystyle \frac{\partial^2 \Psi^i}{\partial x_1 \partial x_1} & 
    \displaystyle \frac{\partial^2 \Psi^i}{\partial x_1 \partial x_2}\\
    &\\
    \displaystyle \frac{\partial^2 \Psi^i}{\partial x_2 \partial x_1} & 
    \displaystyle \frac{\partial^2 \Psi^i}{\partial x_2 \partial x_2}\\
    \end{pmatrix}.
\end{equation}
Finally, we obtain the weak lensing convergence field, $\kappa_s(x_1,x_2)$, from $A_s(x_1,x_2)$ by using the relation of eq.(\ref{eq:kappa_gamma_jacobian}). 
This is the convergence field for one source plane at $\chi_s$.

The convergence map for sources with a redshift distribution of $p(z)$, $\kappa^{\rm tot}$, is obtained by summing $\kappa_s$ with a weight,
\begin{equation}
    \kappa^{\rm tot} = \sum_{i=i_{\rm min}}^{i_{\rm max}}
    		w_i
            \kappa_i,
\end{equation}
with
\begin{equation}
\label{eq:zs_weight}
    w_i = \int_{z_{i}-\Delta z_i/2}^{z_{i}+\Delta z_i/2}{\rm d}z\;p(z).
\end{equation}
Here $z_i$ and $\Delta z_i$ are the redshift and redshift interval of the $i$-th box, and $i_{\rm min}$ and $i_{\rm max}$ are the minimum and maximum redshifts of source galaxies, respectively.
The redshift distribution $p(z)$ should be normalized to unity in $(z_{i_{\rm min}}-\Delta z_{i_{\rm min}}/2) < z < (z_{i_{\rm max}}+\Delta z_{i_{\rm max}}/2)$.
By varying the redshift range of source galaxies (i.e., varying $i_{\rm min}$ and $i_{\rm max}$), we can perform the tomographic analysis of the convergence field.

\begin{figure}[]
  \begin{center}
    \includegraphics[width=\textwidth]{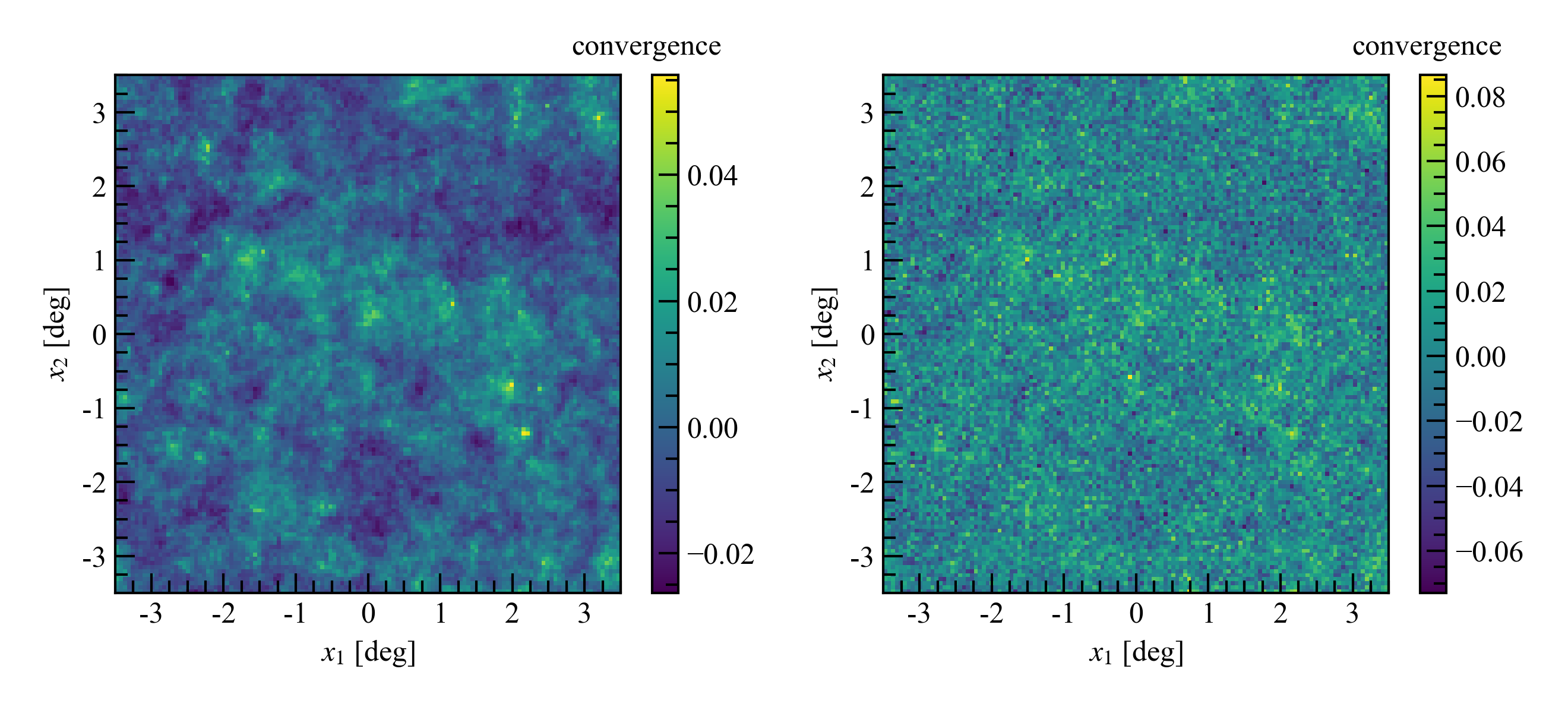}
    \caption{A mock convergence map without (left) and with (right) shape noise. 
    The source galaxies are distributed in $0.6 < z < 3.2$. The shape noise parameters are set to be typical values of the HSC survey,
$\sigma_\gamma=0.22$ and $\bar{n}_g=14.6$ arcmin$^{-2}$.}
    \label{fig:lognormal_lens_map}
  \end{center}
\end{figure}

In the current cosmological weak lensing surveys, statistical errors are dominated by the shape noise from the scatter of intrinsic morphology of source galaxies.
We assume that the shape noise follows a Gaussian distribution with the variance $\sigma_{N}^2$ given by
\begin{equation}
 \sigma_N^2=\frac{\sigma_\gamma^2}{\bar{n}_g\Omega_{\rm pix}},
\end{equation}
where $\Omega_{\rm pix}$ is the pixel size of the convergence map, $\sigma_\gamma$ is the rms shear due to the intrinsic galaxy ellipticity and $\bar{n}_g$ is the mean number density of source galaxies.
Figure \ref{fig:lognormal_lens_map} shows the final lensing convergence map obtained by our {\tt lognormal\_lens} with and without shape noise.
We set $\sigma_\gamma=0.22$ and $\bar{n}_g=14.6$ arcmin$^{-2}$ in $0.6 < z < 3.2$, which are typical
values of the HSC weak lensing survey.

\subsection{Galaxy positions in redshift space}
The {\tt lognormal\_galaxies} code generates the galaxy positions and the velocity field in addition to the matter density field.
We can measure the three-dimensional power spectrum in redshift space as well as the angular power spectrum of galaxies, which should correlate with the convergence field described in the previous section.

To cross-correlate with the convergence fields, we also use the plane-parallel approximation to project the three-dimensional galaxy density field $\delta_{\rm gal}$ onto the two-dimensional galaxy density field $\delta^{\rm proj}_{\rm gal}$ in the same manner as the matter density field,
\begin{equation}
	\delta^{\rm proj}_{\rm gal}(x_1,x_2) =
	\frac1{N_{{\rm grid},y}}
	\sum_{y_i} \delta_{\rm gal}(x_1,x_2,y_i).
\end{equation}
The spatial position $(x_1,x_2)$ is converted to the angular position as $(\theta_1,\theta_2) = (x_1/\chi,x_2/\chi)$ where $\chi$ is the comoving distance to the center of the simulation box.

\subsection{Power spectrum measurement}
From the mock observable fields $x$ and $y$, we compute the angular power spectrum as 
\begin{equation}
    \ell'^2 \mathcal{C}^{xy}(\ell') = 
    \frac{1}{L^2}
    \left[
    \frac{1}{N_{{\rm mode},b}}
    \sum_{\bm{\ell}}^{\ell \in \ell_b} \ell^2 \delta^x(\bm{\ell})
\delta^{y*}(\bm{\ell})
    \right],
\end{equation}
where $L$ is the side-length of the square-shape simulation field, $\ell_b$ denotes the multipole bins, $N_{{\rm mode},b}$ is the number of modes within the bin, and $\delta^x$ and $\delta^y$ are the Fourier transform of the fields $x$ and $y$.
The bin-averaged multipole $\ell'$ is calculated as
\begin{equation}
    \ell' = 
    \frac{1}{N_{{\rm mode},b}}
    \sum_{\bm{\ell}}^{\ell \in \ell_b} \ell.
\end{equation}
We use the logarithmically equally spaced bins with the minimum multipole of $\ell_{\rm min} = 2\pi/L$ and the bin width of ${\rm d}\ln \ell = 0.3\log_{10}{e}$.
For the Fourier transformation of the fields we use the publicly available library of fast Fourier transform, FFTW \cite{FFTW05}.

\section{Validation of the mocks}
\label{sec:validation}
In this section we validate the internal consistency of our algorithm by comparing the angular auto- and cross-power spectra of the convergence and galaxy density fields from mocks with the input ones.
Although it has been proved that the density field generated by {\tt lognormal\_galaxies} precisely follows the input power spectrum \cite{Agrawal/etal:2017}, some systematics can be introduced during the construction of the  convergence fields and measurement of the power spectra. In what follows we prove that both the measured convergence auto- and convergence-galaxy cross-spectra recover the model power spectra which is calculated from the input matter and galaxy power spectra.

\subsection{Simulation settings}
For the fiducial simulation we use a flat $\Lambda$CDM model with the parameters of {\it Planck} 2015 `TT,TE,EE+lowP': $\Omega_b h^2 = 0.02225$, $\Omega_c h^2 = 0.1198$, $n_s = 0.9645$, $\ln(10^{10}A_s) = 3.094$ and $h = 0.67021$ with the minimum neutrino mass of $\sum m_\nu = 0.06$ [eV] \citep{planck2015}.
The input matter power spectra at various redshifts are calculated by the publicly available code {\tt CLASS} \cite{class1, class2}.
For simplicity, we use the linear matter power spectrum.
The default field-of-view of one realization is $7 \times 7$ ${\rm deg}^2$ and the number of two-dimensional angular grids is $N_{\rm grid, 2D} = 256$, unless otherwise noted\footnote{The computational cost for a realization using one CPU core on a laptop PC is a few minutes for the default configuration.}.
The square shape of the map comes from a limitation of RAYTRIX and we may mitigate this limit in near future.
The shape noise is not included, so that we can test the simulation results precisely.

The light cone is generated in $0 < z < 3.2$ as shown in Fig. \ref{fig:lognormal_lens_geometry}, consisting of 15 log-normal boxes with similar sizes along the line of sight.
We confirmed that the box thickness does not significantly affect the resultant angular power spectrum.

Each log-normal box is generated with the three-dimensional grid number of $N_{\rm grid, 3D} = 256$ for the axis perpendicular to the line of sight, while the grid number of the line-of-sight axis, $N_{\rm grid,y}$, is determined so that the physical size of the grid is the same with the other axis.
In the redshift range covered by the PFS cosmology survey, $0.6 < z < 2.4$ (corresponding to the red boxes in Figure \ref{fig:lognormal_lens_geometry}), we also generate galaxy density fields, which will be cross-correlated with the convergence field.

\subsection{Auto-power spectrum of the convergence field}
\subsubsection{Single source redshift}
First, we examine the simplest case: the convergence auto-power spectrum for a single source redshift.
In this limit the lensing kernel $W^{\kappa}$ is simplified to eq.(\ref{eq:wk_single}).

\begin{figure}[]
  \begin{center}
    \includegraphics[width=\textwidth]{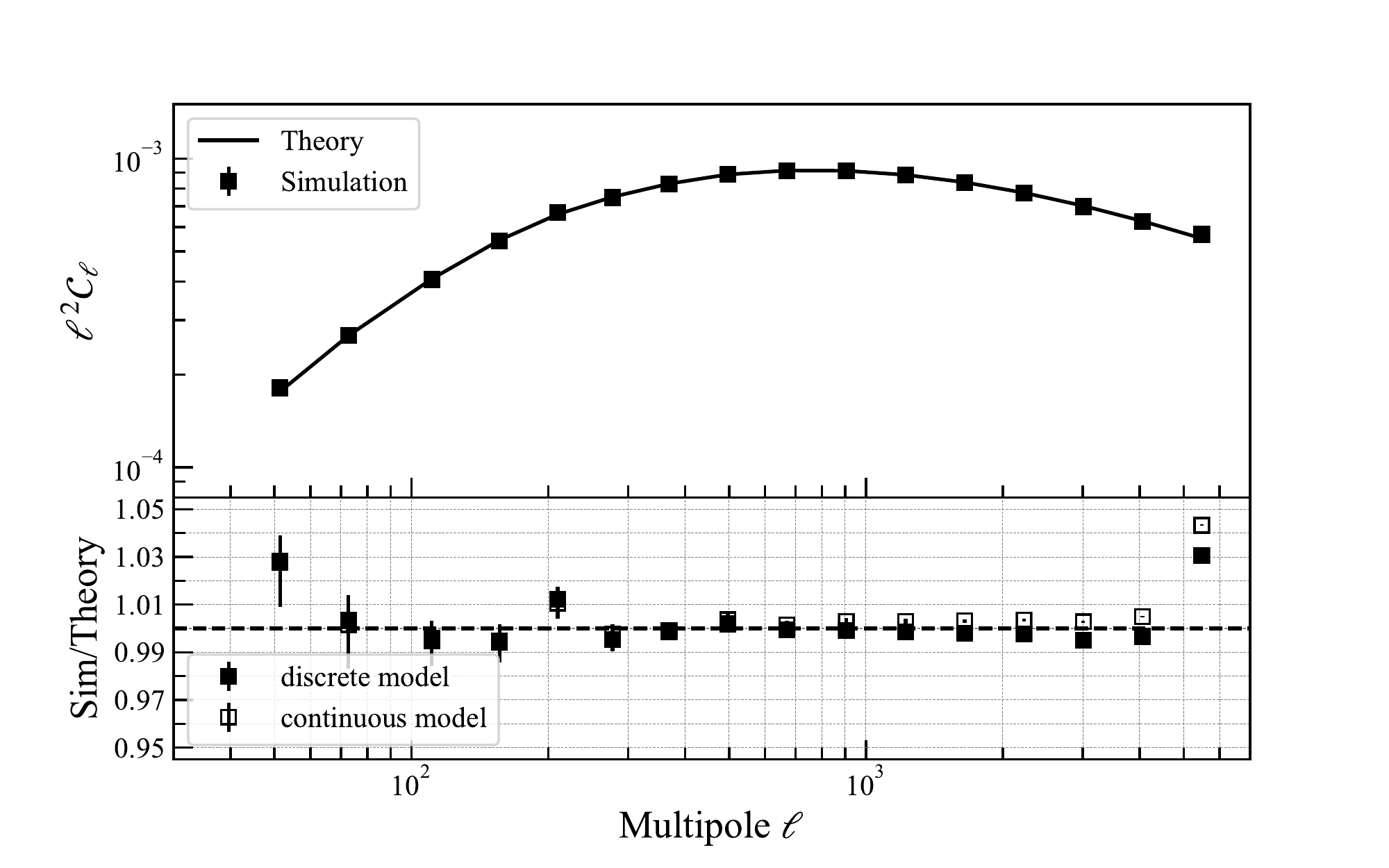}
    \caption{{\it Top}: Mean of 2,000 simulated angular auto power spectra of the convergence field (filled squares). 
    The error of the mean is smaller than the size of the square.
    The single source plane at $z = 2.99$ is assumed.
    The solid line is the theoretical model with the discrete redshift integral.
    { \it Bottom}: The ratio of the simulation and the theoretical model, with (filled) and without (open) discretization of the redshift integral.
    }
    \label{fig:cl_kk_single}
  \end{center}
\end{figure}

The top panel of Figure \ref{fig:cl_kk_single} shows the mock and theoretical auto-power spectra of the convergence field for the source redshift of $z_s = 2.99$.
The simulation data points are the mean of 2,000 realizations, while the error bars show the error of the mean for 68 percentile.

To take into account the effect of discretization in the simulation data, we evaluate the theoretical model on the same grid as in the simulations in multipole space and binned as 
\begin{equation}
    \ell'^2 \mathcal{C}^{\rm theory}(\ell') = 
    \frac{1}{N_{{\rm mode},b}}
    \sum_{\bm{\ell}}^{\ell \in \ell_b} \ell^2 
   	C^{\rm theory}(\ell).
\end{equation}
In the following we use this grid-based theoretical power spectrum when we compare the theory and mock power spectra.

Since the mock light cone is constructed from the discrete boxes, the redshift evolution of the matter density field is not smooth.
To take into account this effect, we also discretize the redshift integration in the theoretical model of the convergence power spectrum, eq.(\ref{eq:clkk}), as
\begin{equation}
	\label{eq:cl_kk_discretized}
    C^{\kappa\kappa}(\ell) = \sum_{i=1}^{n-1}
    \Delta \chi_i
    W_{\kappa}^2(\chi_i)\chi_i^{-2}
    P_{\rm mm}^{i}((\ell+0.5)/\chi_i),
\end{equation}
where $P^i_{\rm mm}$, $\chi_i$ and $\Delta \chi_i$ are the input matter power spectrum, the comoving distance and the thickness of the $i$-th box, respectively. 
The source galaxies are located at $\chi_n$.
The filled squares in the bottom panel of Figure \ref{fig:cl_kk_single} show the ratio of mock and theory obtained by the discretized model, eq.(\ref{eq:cl_kk_discretized}), while the open squares show that with the continuous model, eq.(\ref{eq:clkk}).
The difference between two models is less than $0.5\%$.

The mock power spectrum matches the theoretical model with better than 1\% accuracy, except the largest multipole bin in  which the effect of the finite grid size of the three-dimensional matter density field is significant. 
This trend is also seen in the power spectrum of three-dimensional  density field measured from a single lognormal box, as shown in ref.~\cite{Agrawal/etal:2017}.
Although the lognormal method does not completely capture the real-world physics and therefore less accurate than the first principle simulations, this precise agreement between the simulation output and the input model is the main advantage of this method, as this allows us to test real-world issues without worrying about inaccuracy of the theoretical model for, e.g., $N$-body simulations.

The authors of ref.~\cite{takahashi/etal:2017} show that one needs to take into account the effect of finite thickness of a mass sheet (called ``shell'' in ref.~\cite{takahashi/etal:2017}), because the power in the line-of-site direction on scales larger than the mass sheet thickness is suppressed.
In our simulation, however, this effect does not appear. 
We generate a mass sheet at a given redshift by projecting the entire simulation box at that redshift, which satisfies the periodic boundary condition; thus, the large scale mode is not lost. 
On the other hand, the ``shell'' in ref.\cite{takahashi/etal:2017} is cut out from the simulation box larger than the shell thickness and therefore the window effect is induced.

As noted above we adopt the plane-parallel approximation when we calculate paths of light rays at the position of each mass sheet.
This approximation is valid when the opening angle of the survey is sufficiently small.
Although our simulation outputs would match the theoretical prediction from eq.(\ref{eq:cl_kk_discretized}) with arbitrary opening angles by design, such a simulation does not capture the effect of curvature of sky.
We can try to incorporate the effect of sky curvature by radially propagating light rays from the observer position. We relate the angular position $\bm{\theta}^i = (\theta_1,\theta_2)$ and the spatial position $\bm{x}^i=(x_1,x_2)$  as $(\tan{\theta_1},\tan{\theta_2}) = (x_1/\chi_i,x_2/\chi_i)$.
In this case, however, the angular grids and the spatial grids no longer match; thus, we use the cloud-in-cell (CIC) interpolation to estimate the deflection potential at arbitrary angular positions.

\begin{figure}[]
  \begin{center}
    \includegraphics[width=\textwidth]{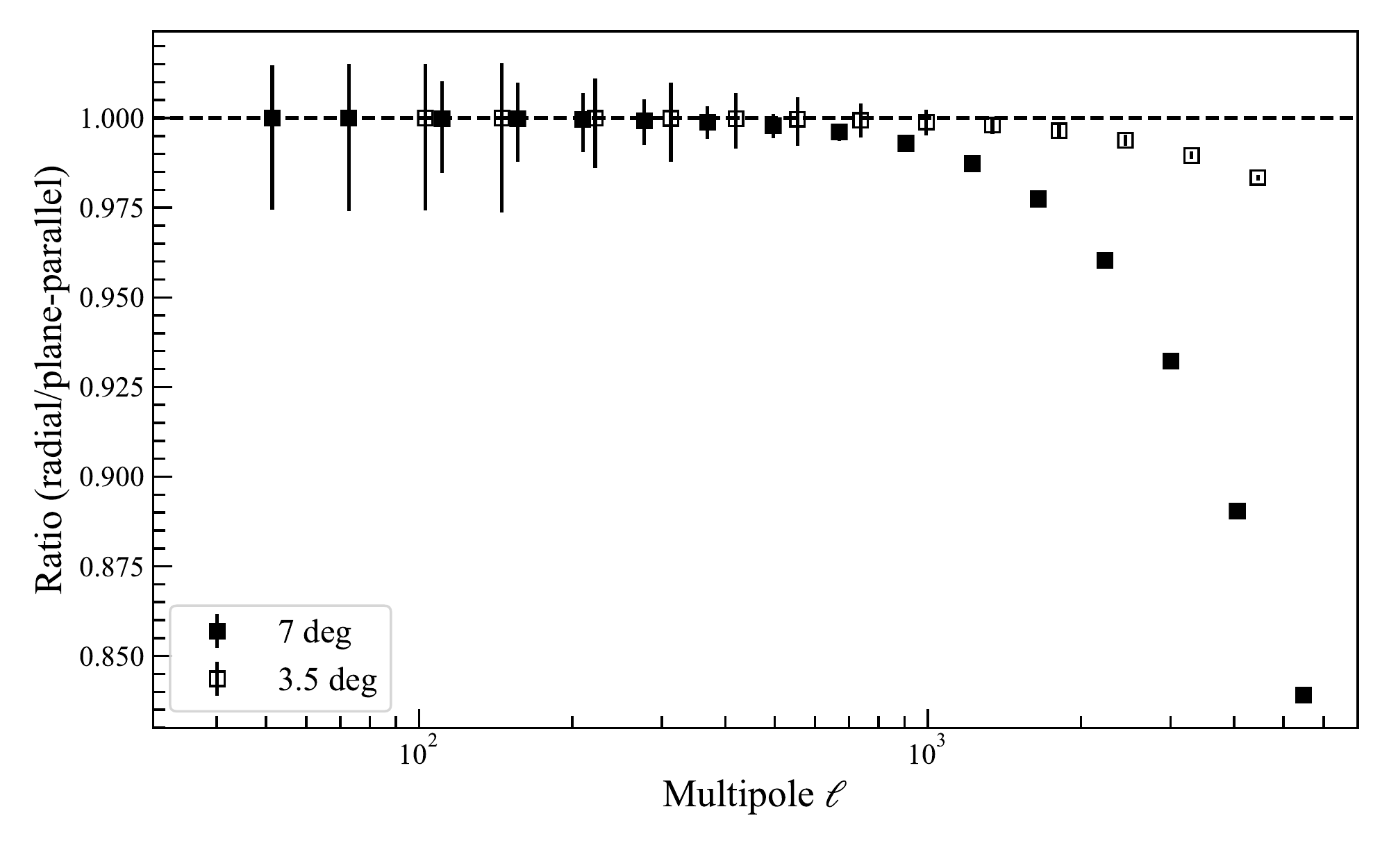}
    \caption{The ratio of the mock convergence auto-power spectra with the ``radial ray propagation'' and ``plane-parallel ray propagation'' methods. The filled squares show the results of $7\times7\;{\rm deg}^2$ and $N_{\rm grid, 2D} = 256$ simulation, while the open squares show the $3.5\times3.5\;{\rm deg}^2$ and $N_{\rm grid, 2D} = 128$ simulation.
    }
    \label{fig:cl_kk_radial}
  \end{center}
\end{figure}

Figure \ref{fig:cl_kk_radial} shows the ratio of the ``radial ray propagation'' to the ``plane-parallel ray propagation'' (fiducial model) results for opening angles of $7\times7\;{\rm deg}^2$ and $3.5\times3.5\;{\rm deg}^2$. 
The number of grids for the $3.5$ deg simulation is 128, which is chosen to yield the same grid size as for the $7$ deg simulation, i.e., both simulations have the same spatial resolution. 
We find that the small-scale power is suppressed in the ``radial ray propagation'' case.
This is because the CIC interpolation smears out the small scale fluctuations.
The suppression is more significant for the $7$ deg simulation, because the difference in the paths of light rays between the ``radial ray propagation'' and ``plane-parallel ray propagation'' cases becomes larger at larger angles.
This result shows that the projection of the density field and light-ray propagation should be done in the same way.
The angular coordinate version of the RAYTRIX, GRayTrix \cite{shirasaki/etal:2015,hamana/etal:2015}, can be used to perform the large opening angle simulation. Implementation of this is left for future work.

\subsubsection{Multiple source redshifts}
\begin{figure}[]
  \begin{center}
    \includegraphics[width=\textwidth]{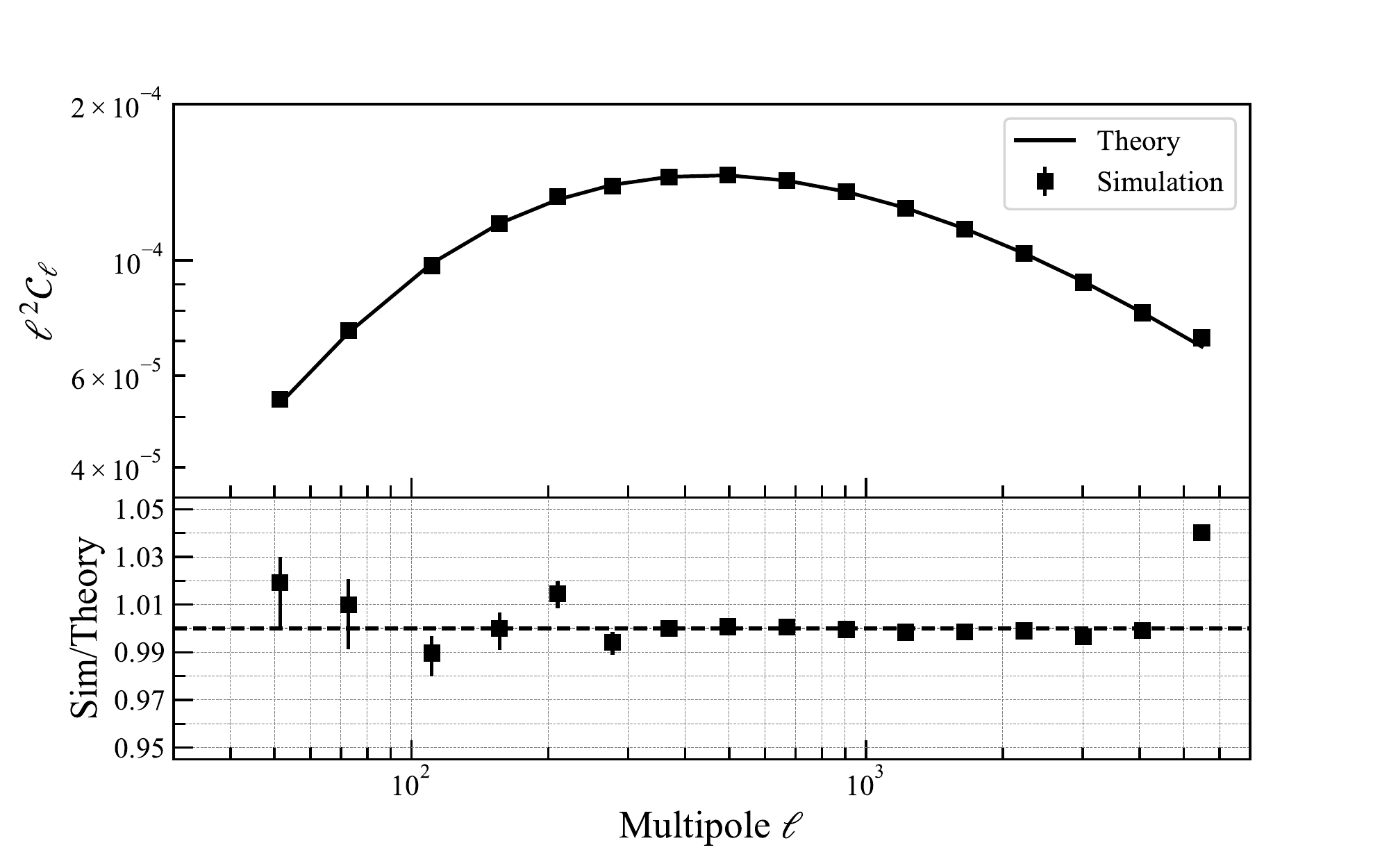}
    \caption{
    Same as Figure~\ref{fig:cl_kk_single}, but for multiple source planes at $0.6 < z < 3.2$.
    }
    \label{fig:cl_kk_multi}
  \end{center}
\end{figure}

Next we investigate the convergence auto power spectrum with multiple source planes.
In this case the lensing kernel $W_{\kappa}$ is given in eq.(\ref{eq:wk_multi}).
Since the redshift distribution of source planes is discrete in our  simulation, we also discretize the redshift integral in eq.(\ref{eq:wk_multi}) as
\begin{equation}
    W_\kappa(\chi_i) = 
    \frac{3H_0^2\Omega_m}{2c^2}
    \sum_{j=i}^{n-1}
    \Delta \chi_j
    w_{j}\frac{\chi_{i}}{a(\chi_{i})}
    \left(1-\frac{\chi_{i}}{\chi_{j}}\right),
\end{equation}
where $w_i$ is the weight defined in eq.(\ref{eq:zs_weight}).
For the redshift distribution of source galaxies, $p(z)$, we assume 
\begin{equation}
	p(z) = z^2 \exp(-z/z_0), 
\end{equation}
with $z_0 = 1/3$, which approximates the redshift distribution of source galaxies in the HSC weak lensing survey.
Figure \ref{fig:cl_kk_multi} shows the convergence auto-power spectrum with source galaxies at $z > 0.6$.
The mock spectrum matches the theoretical prediction at the same level as in the single source redshift case shown in Figure \ref{fig:cl_kk_single}.

\subsection{Cross spectra}
\begin{figure}[]
  \begin{center}
    \includegraphics[width=\textwidth]{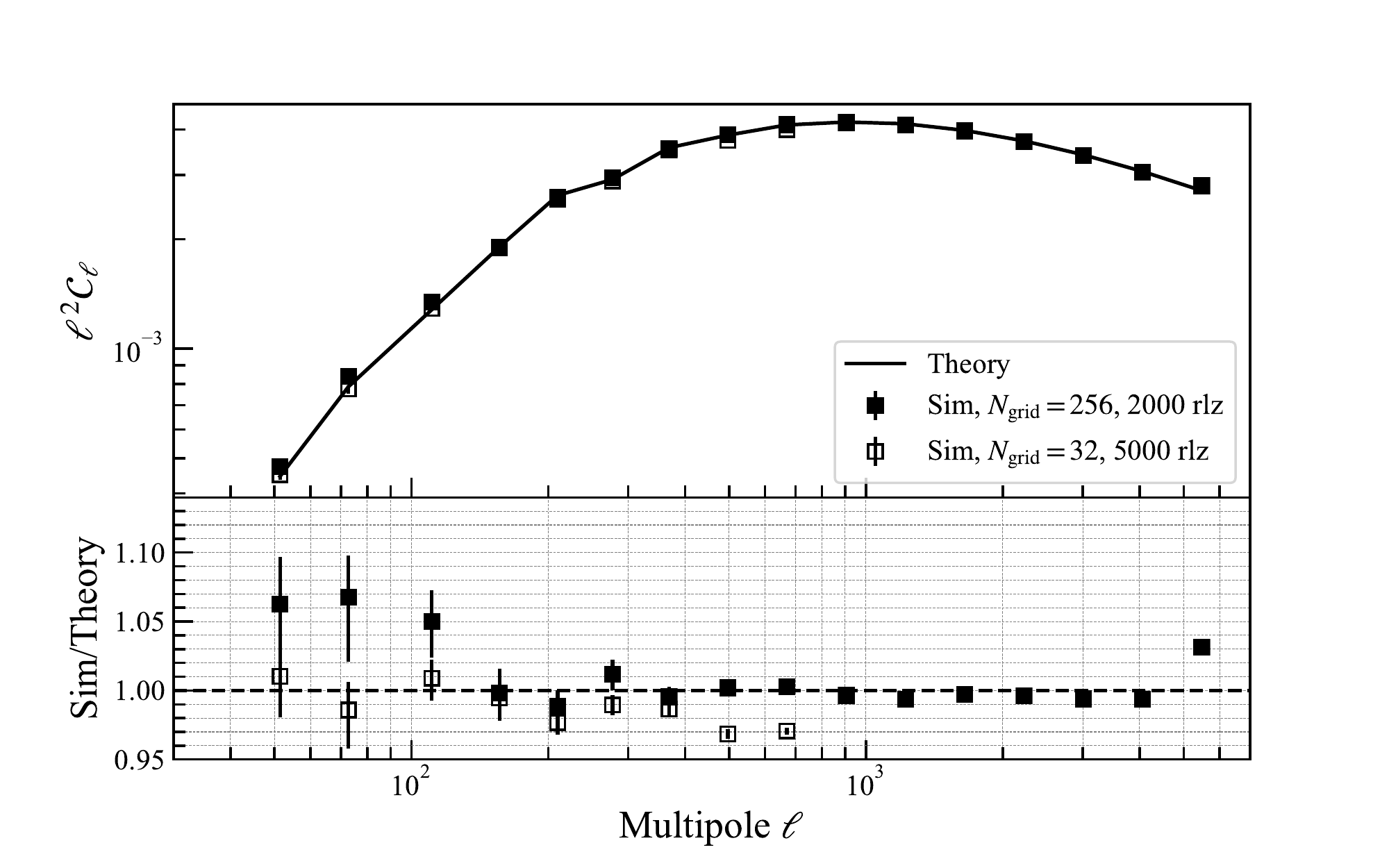}
    \caption{{ \it Top}: 
    Mean of 2,000 simulated galaxy-convergence cross-power spectra for galaxies in the redshift slice of $1.2<z<1.4$ and the single source redshift of $z_s = 2.99$
    (filled squares). The error of the mean is smaller than the size of the square.
    The white squares show the mean of 5,000 realizations with $N_{\rm grid, 2D}=32$, to show less noisy results at small multipoles. The black solid line is the theoretical model. { \it Bottom}: The ratio of the simulations and the model.
    }
    \label{fig:cl_gk_single}
  \end{center}
\end{figure}

\begin{figure}[]
  \begin{center}
    \includegraphics[width=\textwidth]{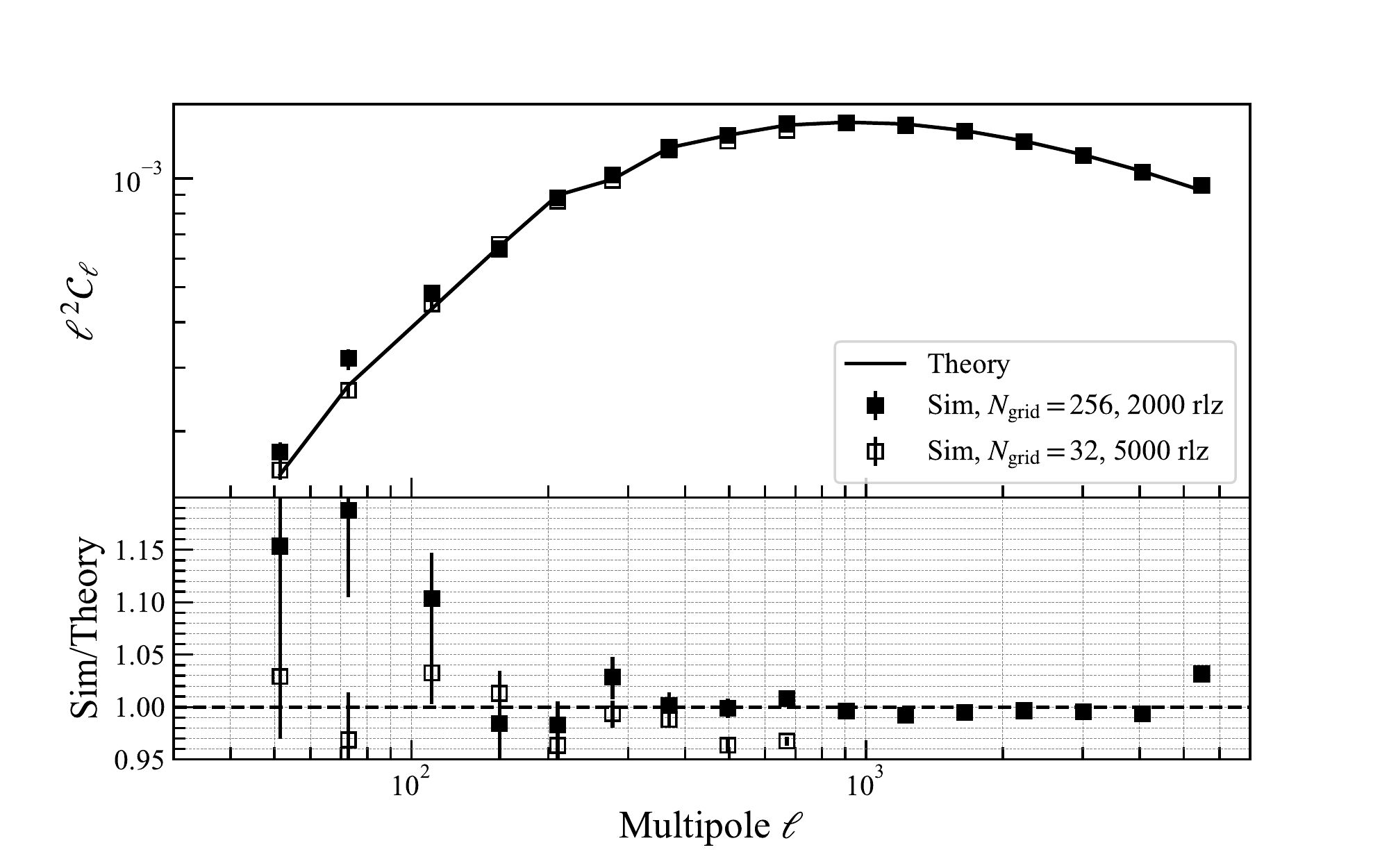}
    \caption{
    Same as Figure~\ref{fig:cl_gk_single}, but for multiple source planes at $1.4 < z < 3.2$.
    }
    \label{fig:cl_gk_multi}
  \end{center}
\end{figure}

Figure \ref{fig:cl_gk_single} and \ref{fig:cl_gk_multi} show the mock galaxy-convergence cross-power spectrum for a single source plane at $z_s = 2.99$ and multiple source planes at $1.4 < z < 3.2$, respectively. 
The galaxies are located in the redshift slice of $1.2<z<1.4$, with the galaxy bias and number density set to be $b_g = 1.42$ and $n_g = 7.8\times10^{-2}$ $h^3\; {\rm Mpc}^{-3}$ assuming the PFS cosmology survey (Table~\ref{tb:pfs}) but 100 times higher number density, to suppress the shot noise.
The data points are the mean of 2,000 realizations, while the error bars show the error of the mean for 68 percentile.

As with the lensing auto-power spectrum, the theoretical model of the cross spectrum defined in eq.(\ref{eq:clgk}) is also modified as
\begin{equation}
    C^{g\kappa}(\ell) = 
    \Delta \chi_i
    W_g(\chi_i) W_{\kappa}(\chi_i)\chi_i^{-2}
    P_{\rm gm}^{i}((\ell+0.5)/\chi_i),
\end{equation}
where the cross correlation is taken with galaxy samples of the $i$-th box.
The galaxy kernel $W_g$ is the normalized redshift distribution of spectroscopic galaxy samples. 
Since we assume a constant number density within each box, $W_g = 1/\Delta \chi_i$.
The galaxy-matter cross power spectrum $P_{\rm gm}$ is calculated from the input matter and galaxy power spectra following the procedure of ref.~\cite{Agrawal/etal:2017}.

We find that the mock cross power spectra match the theoretical ones with 1\% accuracy except the highest multipole, although the statistical error is large at lower multipoles. 
To asses potential systematics at lower multipoles, we further generate 5,000 realizations with coarser grids, $N_{\rm grid, 2D} =32$.
The results are shown in the open squares in  Figure \ref{fig:cl_gk_single} and \ref{fig:cl_gk_multi}.
We find that there is no systematic deviation from the theoretical predictions at low multipoles.

\begin{figure}[]
  \begin{center}
    \includegraphics[width=\textwidth]{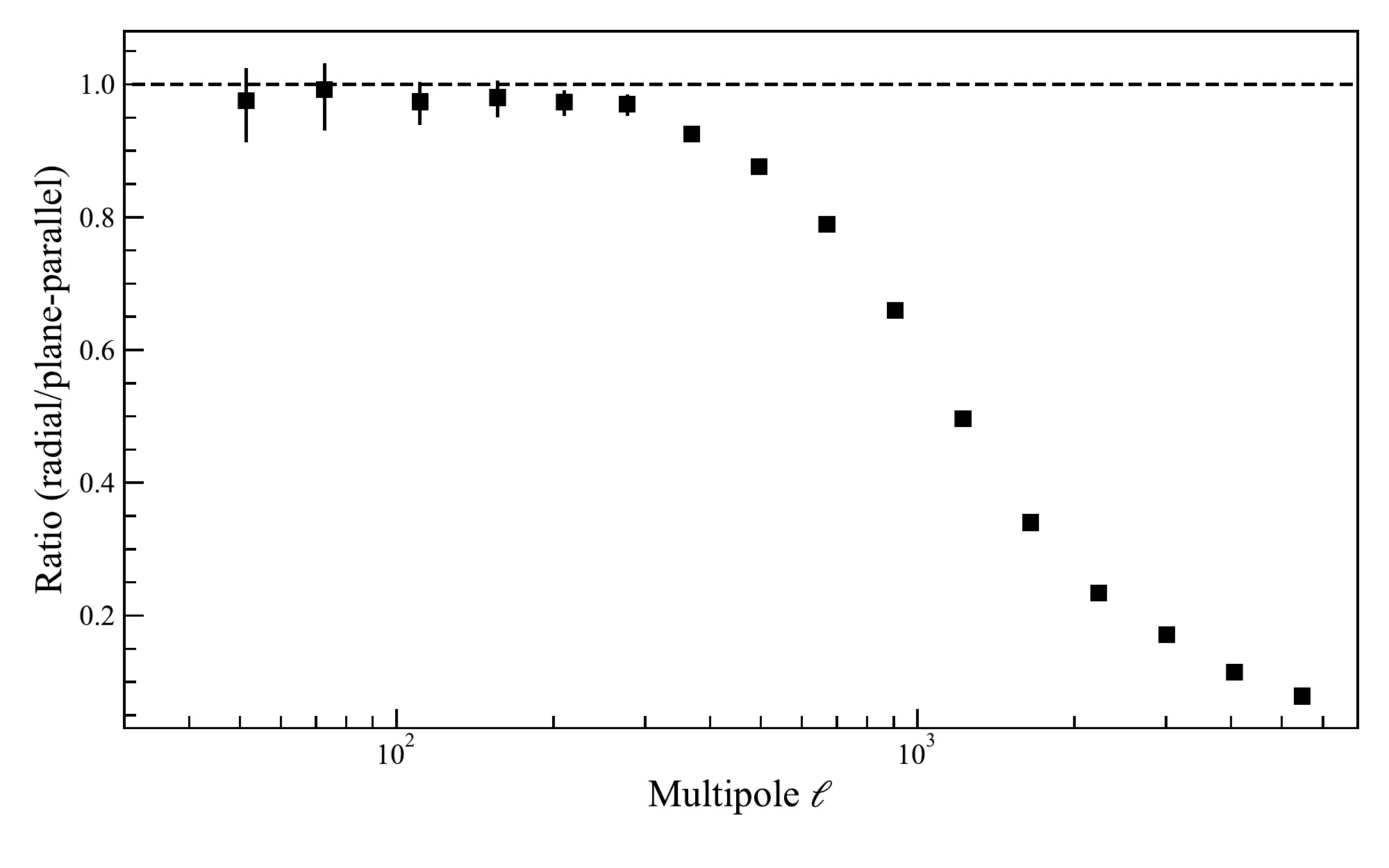}
    \caption{
The ratio of the mock galaxy-convergence cross-power spectra with the “radial galaxy projection” and  “plane-parallel galaxy projection” methods.
    }
    \label{fig:cl_gk_radial}
  \end{center}
\end{figure}

As noted above, the projection of matter
and galaxy density fields as well as the propagation of light rays are all done by the plane-parallel approximation.
If one radially projects galaxies into a two-dimensional map, it would suppress the small-scale power due to mismatch of the galaxy positions and the underlying matter density grids. Figure \ref{fig:cl_gk_radial} shows the ratio of the cross-power spectrum estimated from the radially projected galaxy map to that from the parallel projection.
We find that the small scale power is significantly suppressed for the former case, showing again importance of using the same projection method for all the quantities involved. 

\section{Forecasting the Subaru PFS and HSC}
\label{sec:forecast}
The PFS cosmology survey will be conducted on top of the photometric galaxy catalog obtained by the HSC survey, which provides weak lensing cosmic shear maps through accurate measurements of galaxy shapes.
The combination of the PFS and HSC data provides galaxy-shear power spectra, galaxy power spectra in redshift space, and
cosmic shear power spectra  over a wide range of redshifts.

We perform the {\tt lognormal\_lens} simulations for the PFS and HSC survey parameters.
The survey volume, galaxy number density and  galaxy bias of the PFS cosmology survey are summarized in Table \ref{tb:pfs}.
For the HSC survey, we assume that source galaxies have the shape noise parameter of $\sigma_{\gamma} = 0.22$, the number density of $n_{g, {\rm source}} = 20\;{\rm arcmin}^{-2}$ in $0.0 < z < 3.2$ and the redshift distribution of $p(z) = z^2\exp{(-z/z_0)}$ with $z_0 = 1/3$.

The total survey area of the HSC and PFS survey is about $1400\;{\rm deg}^2$.
To save computational costs (dominated by the high spatial resolution required for the lensing simulation), we generate 30 realizations of $7\times7\;{\rm deg}^2$ fields with $N_{\rm grid, 2D} = 128$, and average them to mock power spectra measured from the total survey area of $7\times7\times30 = 1470\;{\rm deg}^2$.
To measure the covariance matrix we generate  200 realizations of the mock (i.e., 6000 realizations of the $7\times7\;{\rm deg}^2$ simulation in total).

\begin{figure}[]
    \centering
    \includegraphics[width=\textwidth, bb = 0 0 747
562]{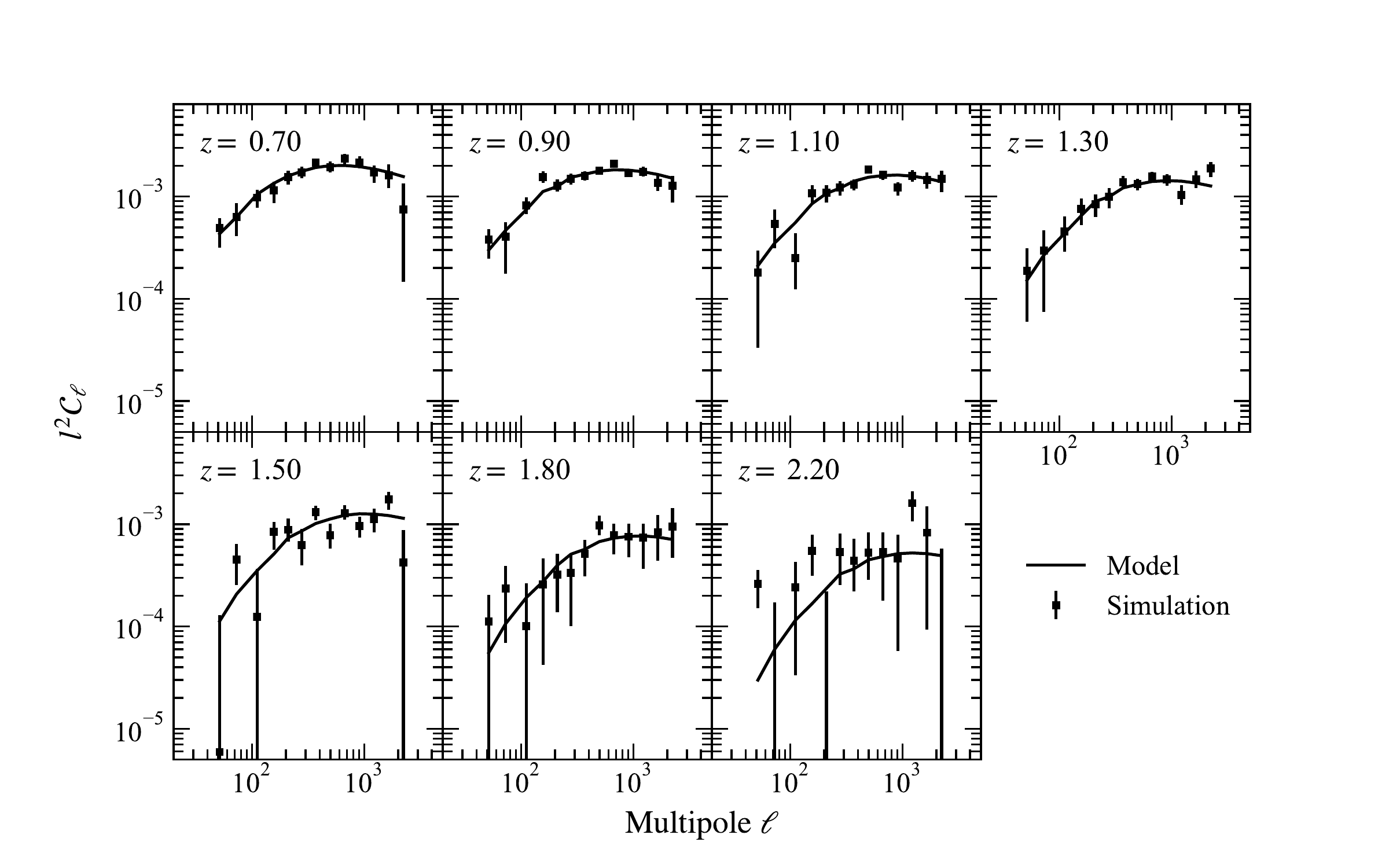}
    \caption{Galaxy-convergence cross power spectra at seven tomographic redshift bins, expected from the HSC and PFS surveys. 
    The data points show one realization of the simulation. 
    The error bars are the diagonal terms of the covariance matrix estimated from 200 realizations.
    The solid black lines show the theoretical model.}
    \label{fig:cl_gk_pfs}
\end{figure}

Figure \ref{fig:cl_gk_pfs} shows the galaxy-convergence cross-power spectra at 7 tomographic redshift bins. 
We only use the source galaxies that are located at higher redshifts than the corresponding PFS galaxy samples. 
This figure provides a visual representation of the quality of galaxy-shear power spectra expected from the HSC and PFS surveys. 

\begin{figure}[]
    \centering
    \includegraphics[width=\textwidth, bb = 0 0 576 461]{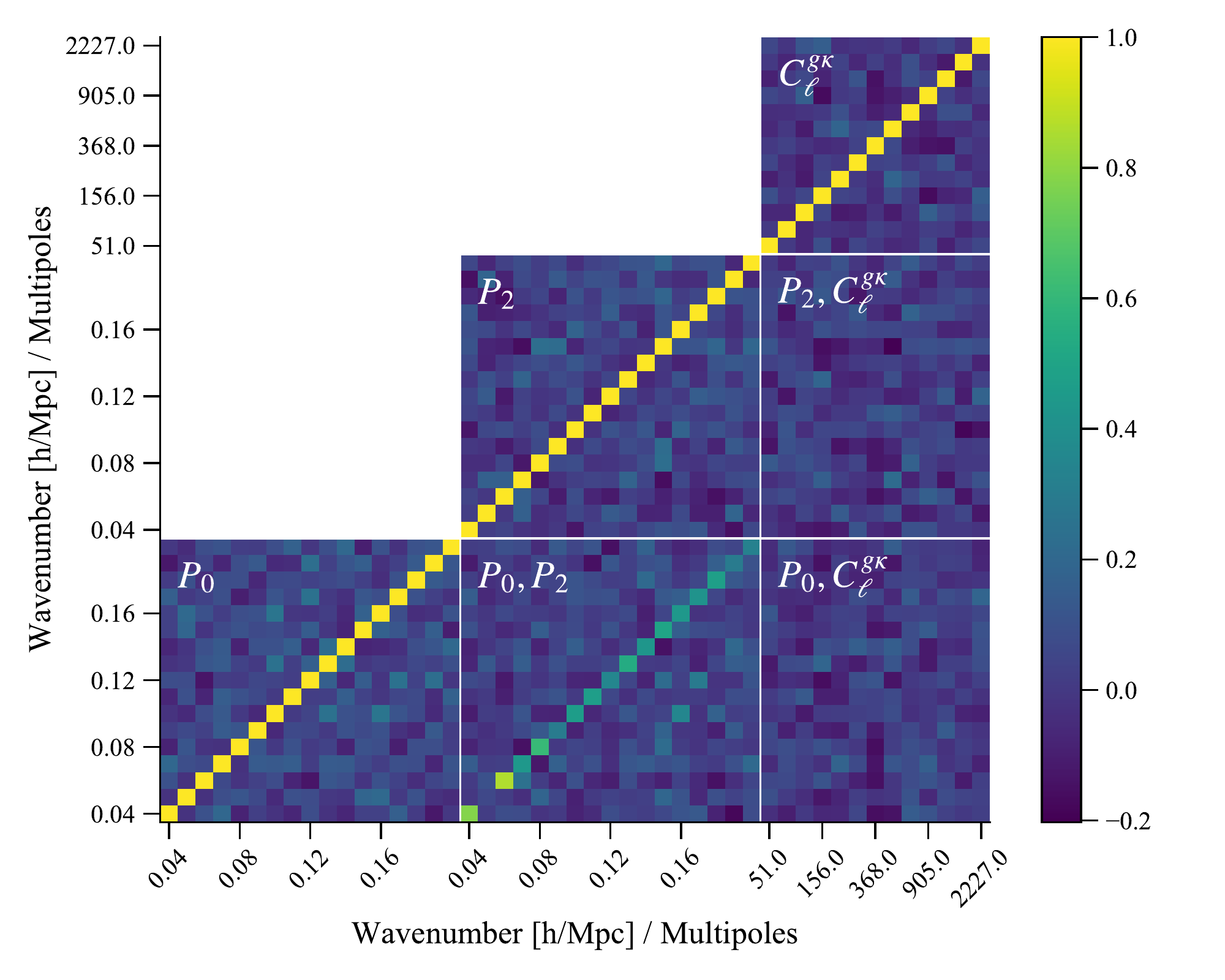}
    \caption{Correlation coefficient matrix for the galaxy power spectrum
monopole ($P_0$), quadrupole ($P_2$) and the galaxy-convergence cross-power
spectrum ($C_{\ell}^{g\kappa}$) at $1.2 < z < 1.4$ with source galaxies at $1.4 < z < 3.2$, estimated from 200 realizations of the mock.}
    \label{fig:cov_pfs}
\end{figure}

Figure \ref{fig:cov_pfs} shows the correlation coefficient matrix of the galaxy power spectrum multipoles and the galaxy-convergence cross spectra at $1.2 < z < 1.4$, estimated from 200 realizations of the mock. 
The range of wave numbers shown in this figure, $0.04 < k < 0.20~h/{\rm Mpc}$, roughly corresponds to the multipole range of $110 < \ell < 550$ at $z = 1.3$.
Our simulation results show that the covariance between the galaxy power spectrum multipoles and the galaxy-convergence cross spectrum is negligible.

We use these covariance matrices to calculate the signal-to-noise ratios of the cross spectra as
\begin{equation}
	\left(\frac{S}{N}\right)^2 = \sum_{\ell\le \ell'}{\bm C}_{{\ell}}\;{\rm
Cov}^{-1}(\ell,\ell')\;{\bm C}_{\ell'}{}^{\rm T}.
\end{equation}
We find that the PFS and HSC can measure the cross spectra with signal-to-noise ratios of 20.7, 27.6, 20.6, 18.6, 11.8, 7.6 and 5.0 at redshift bins of $z= 0.7, 0.9, 1.1, 1.3, 1.5, 1.8$ and 2.2 ($50 < \ell < 1000$), respectively. 
It is remarkable that secure detection of the cross power spectra is expected out to such high $z$, showing the power of simultaneous imaging and spectroscopy using a 8-m class telescope.

For the purpose of the validation of mock catalogs, we also calculate the signal-to-noise ratios analytically by using the following expression for the Gaussian term of the covariance matrix \cite{hu/jain:2004},
\begin{equation}
{\rm Cov}^{\rm G}(C_{\ell_1}^{g\kappa},C_{\ell_2}^{g\kappa})
= \frac{\delta_{\ell_1 \ell_2}}{f_{\rm sky}(2\ell_1+1)\Delta \ell_1}
\left[
{C}_{\ell_1}^{g\kappa}{C}_{\ell_2}^{g\kappa}+
\hat{C}_{\ell_1}^{gg}\hat{C}_{\ell_2}^{\kappa \kappa}
\right],	
\end{equation}
where $\delta$ is the Kronecker delta, $\Delta \ell$ is the multipole bin size, and $f_{\rm sky} = 0.036$ $(1470\;{\rm deg}^2)$ is the available sky fraction of the assumed survey.
The galaxy and convergence auto-power spectra with hat symbol, $\hat{C}_{\ell_1}^{gg}$ and $\hat{C}_{\ell_2}^{\kappa \kappa}$, include the shot noise and shape noise term, respectively.
They are written as
\begin{eqnarray}
	\hat{C}_{\ell}^{gg} &=& C_\ell^{gg}+\frac{1}{\bar{n}_{\rm g,spec}(z)}, \\
	\hat{C}_{\ell_2}^{\kappa \kappa} &=& C_\ell^{\kappa \kappa}+\frac{\sigma_{\gamma}^2}{\bar{n}_{\rm g,source}(z)},
\end{eqnarray}
where the $\bar{n}_{\rm g,spec}(z)$ and $\bar{n}_{\rm g,lens}(z)$ are the galaxy number density per steradian of the spectroscopic samples and source galaxies at each tomographic redshift bin, respectively.
We obtain the signal-to-noise ratios of 20.8, 25.4, 19.7, 16.3, 11.4, 7.0 and 3.3 at $z = 0.7, 0.9, 1.1, 1.3, 1.5, 1.8$ and $2.2$ $(50 < \ell < 1000)$, which are similar to that obtained from the simulation.

\section{Summary and Conclusions}
\label{sec:summary}
We have presented the new simulation code to generate the weak lensing field based on the log-normal method of generating three-dimensional matter density, velocity and galaxy density fields. 
Ray-tracing the matter density fields, the code self-consistently provides the weak lensing auto power spectra, the galaxy-lensing cross-spectra and the galaxy power spectra in redshift space.
The code thus offers a useful tool for generating mock observations of on-going and future LSS surveys.

The mock power spectra agree with the input model with better than 1\% accuracy, which is the main advantage of the log-normal method. 
To achieve this precision, we found that the subtle systematics is introduced if we ignore the discreteness of simulation boxes when calculating the theoretical power spectrum.
We also found that the projection of matter density and galaxy density fields and the propagation of light rays must be done in the consistent manner; otherwise, the small-scale power would be significantly suppressed.

Using this new code, we have presented forecasts for the future cosmology survey of Subaru HSC and PFS. 
We found that the combination of HSC and PFS data would detect the galaxy-lensing cross spectra with high signal-to-noise ratios out to unprecedentedly high redshifts.
We also found that the cross-covariance between the galaxy power spectrum multipoles and the galaxy-lensing cross-spectra can be ignored at the level of statistical uncertainties of the HSC and PFS surveys.

One of the advantages of the log-normal method is that the outputs are predictable from the inputs with high accuracy, as validated in this paper.
In future we use the simulation to test several systematics that arise in the real observations, e.g., the uncertainties of photometric redshift of source galaxies, non-uniform survey completeness, fiber collision, and so on.

\acknowledgments
We thank Aniket Agrawal, Aoife Boyle, Shun Saito, and Keitaro Takahashi for discussions. 
This work was supported in part by JSPS  KAKENHI  Grant Number JP15H05896 (RM, EK, IK), JP20K14515 (RM), and JP20K04016 (IK).
The Kavli IPMU is supported by World Premier International Research Center Initiative (WPI), MEXT, Japan.

\appendix
\section{Modifications to RAYTRIX}
\label{sec:code_docs}
To combine with {\tt lognormal\_galaxies}, we modify RAYTRIX as follows: 
a) The original code requires the depth of all
mass sheets to be the same (in the comoving scale). We loosen this requirement so that the code is able to handle mass sheets with various depth, because we have the PFS survey, for example, in mind, where the depth of the planned survey geometry differs between redshift ranges. 
b) We add a function to generate a lensing map for source galaxies with broad redshift distribution, although the original code provides maps for source galaxies on given redshifts. 
c) We may also add shape noise on the resulting convergence field with the simple Gaussian distribution. 
We package the source codes of modified RAYTRIX into {\tt lognormal\_lens} and make it public on the internet\footnote{See \url{http://wwwmpa.mpa-garching.mpg.de/~komatsu/codes.html}}.

\bibliography{main}

\providecommand{\href}[2]{#2}\begingroup\raggedright\begin{thebibliography}{10}

\bibitem{peebles:1980}
P.~J.~E. {Peebles}, \emph{{The large-scale structure of the universe}}.
\newblock Princeton University Press, 1980.

\bibitem{alam/etal:2017}
{BOSS Collaboration}, S.~{Alam}, M.~{Ata}, S.~{Bailey}, F.~{Beutler},
  D.~{Bizyaev} et~al., \emph{{The clustering of galaxies in the completed
  SDSS-III Baryon Oscillation Spectroscopic Survey: cosmological analysis of
  the DR12 galaxy sample}},
  \href{https://doi.org/10.1093/mnras/stx721}{\emph{Mon. Not. R. Astron. Soc.}
  {\bfseries 470} (Sept., 2017) 2617--2652},
  [\href{https://arxiv.org/abs/1607.03155}{{\ttfamily 1607.03155}}].

\bibitem{alam/etal:2020}
{eBOSS Collaboration}, S.~{Alam}, M.~{Aubert}, S.~{Avila}, C.~{Balland}, J.~E.
  {Bautista} et~al., \emph{{The Completed SDSS-IV extended Baryon Oscillation
  Spectroscopic Survey: Cosmological Implications from two Decades of
  Spectroscopic Surveys at the Apache Point observatory}}, {\emph{arXiv
  e-prints} (July, 2020) arXiv:2007.08991},
  [\href{https://arxiv.org/abs/2007.08991}{{\ttfamily 2007.08991}}].

\bibitem{troxel/etal:2018}
M.~A. {Troxel}, N.~{MacCrann}, J.~{Zuntz}, T.~F. {Eifler}, E.~{Krause},
  S.~{Dodelson} et~al., \emph{{Dark Energy Survey Year 1 results: Cosmological
  constraints from cosmic shear}},
  \href{https://doi.org/10.1103/PhysRevD.98.043528}{\emph{Phys. Rev. D}
  {\bfseries 98} (Aug., 2018) 043528},
  [\href{https://arxiv.org/abs/1708.01538}{{\ttfamily 1708.01538}}].

\bibitem{hikage/etal:2019}
C.~{Hikage}, M.~{Oguri}, T.~{Hamana}, S.~{More}, R.~{Mandelbaum}, M.~{Takada}
  et~al., \emph{{Cosmology from cosmic shear power spectra with Subaru Hyper
  Suprime-Cam first-year data}},
  \href{https://doi.org/10.1093/pasj/psz010}{\emph{Publ. Astron. Soc. Jap.}
  {\bfseries 71} (Apr., 2019) 43},
  [\href{https://arxiv.org/abs/1809.09148}{{\ttfamily 1809.09148}}].

\bibitem{hildebrandt/etal:2020}
H.~{Hildebrandt}, F.~{K{\"o}hlinger}, J.~L. {van den Busch}, B.~{Joachimi},
  C.~{Heymans}, A.~{Kannawadi} et~al., \emph{{KiDS+VIKING-450: Cosmic shear
  tomography with optical and infrared data}},
  \href{https://doi.org/10.1051/0004-6361/201834878}{\emph{Astron. Astrophys.}
  {\bfseries 633} (Jan., 2020) A69},
  [\href{https://arxiv.org/abs/1812.06076}{{\ttfamily 1812.06076}}].

\bibitem{heymans/etal:2020}
C.~{Heymans}, T.~{Tr{\"o}ster}, M.~{Asgari}, C.~{Blake}, H.~{Hildebrandt},
  B.~{Joachimi} et~al., \emph{{KiDS-1000 Cosmology: Multi-probe weak
  gravitational lensing and spectroscopic galaxy clustering constraints}},
  {\emph{arXiv e-prints} (July, 2020) arXiv:2007.15632},
  [\href{https://arxiv.org/abs/2007.15632}{{\ttfamily 2007.15632}}].

\bibitem{eisenstein/etal:2005}
{\scshape SDSS} collaboration, D.~J. Eisenstein et~al., \emph{{Detection of the
  Baryon Acoustic Peak in the Large-Scale Correlation Function of SDSS Luminous
  Red Galaxies}}, \href{https://doi.org/10.1086/466512}{\emph{Astrophys. J.}
  {\bfseries 633} (2005) 560--574},
  [\href{https://arxiv.org/abs/astro-ph/0501171}{{\ttfamily
  astro-ph/0501171}}].

\bibitem{cole/etal:2005}
{\scshape 2dFGRS} collaboration, S.~Cole et~al., \emph{{The 2dF Galaxy Redshift
  Survey: Power-spectrum analysis of the final dataset and cosmological
  implications}},
  \href{https://doi.org/10.1111/j.1365-2966.2005.09318.x}{\emph{Mon. Not. Roy.
  Astron. Soc.} {\bfseries 362} (2005) 505--534},
  [\href{https://arxiv.org/abs/astro-ph/0501174}{{\ttfamily
  astro-ph/0501174}}].

\bibitem{jackson:1972}
J.~C. {Jackson}, \emph{{A critique of Rees's theory of primordial gravitational
  radiation}}, \href{https://doi.org/10.1093/mnras/156.1.1P}{\emph{Mon. Not. R.
  Astron. Soc.} {\bfseries 156} (Jan., 1972) 1P},
  [\href{https://arxiv.org/abs/0810.3908}{{\ttfamily 0810.3908}}].

\bibitem{sargent/turner:1977}
W.~L.~W. {Sargent} and E.~L. {Turner}, \emph{{A statistical method for
  determining the cosmological density parameter from the redshifts of a
  complete sample of galaxies.}},
  \href{https://doi.org/10.1086/182362}{\emph{Astrophys. J. Lett.} {\bfseries
  212} (Feb., 1977) L3--L7}.

\bibitem{kaiser:1987}
N.~Kaiser, \emph{{Clustering in real space and in redshift space}}, {\emph{Mon.
  Not. Roy. Astron. Soc.} {\bfseries 227} (1987) 1--27}.

\bibitem{desjacques/etal:2018}
V.~{Desjacques}, D.~{Jeong} and F.~{Schmidt}, \emph{{Large-scale galaxy bias}},
  \href{https://doi.org/10.1016/j.physrep.2017.12.002}{\emph{Phys. Rept.}
  {\bfseries 733} (Feb., 2018) 1--193},
  [\href{https://arxiv.org/abs/1611.09787}{{\ttfamily 1611.09787}}].

\bibitem{schneider/ehlers/falco:1992}
P.~{Schneider}, J.~{Ehlers} and E.~E. {Falco}, \emph{{Gravitational Lenses}}.
\newblock Berlin: Springer, Verlag, 1992,
  \href{https://doi.org/10.1007/978-3-662-03758-4}{10.1007/978-3-662-03758-4}.

\bibitem{takada/etal:2014}
{\scshape PFS Team} collaboration, M.~Takada et~al., \emph{{Extragalactic
  science, cosmology, and Galactic archaeology with the Subaru Prime Focus
  Spectrograph}}, \href{https://doi.org/10.1093/pasj/pst019}{\emph{Publ.
  Astron. Soc. Jap.} {\bfseries 66} (2014) R1},
  [\href{https://arxiv.org/abs/1206.0737}{{\ttfamily 1206.0737}}].

\bibitem{desi}
M.~{Levi}, C.~{Bebek}, T.~{Beers}, R.~{Blum}, R.~{Cahn}, D.~{Eisenstein}
  et~al., \emph{{The DESI Experiment, a whitepaper for Snowmass 2013}},
  {\emph{arXiv e-prints} (Aug., 2013) arXiv:1308.0847},
  [\href{https://arxiv.org/abs/1308.0847}{{\ttfamily 1308.0847}}].

\bibitem{lsst}
{\v Z}.~{Ivezi{\'c}}, S.~M. {Kahn}, J.~A. {Tyson}, B.~{Abel}, E.~{Acosta},
  R.~{Allsman} et~al., \emph{{LSST: From Science Drivers to Reference Design
  and Anticipated Data Products}},
  \href{https://doi.org/10.3847/1538-4357/ab042c}{\emph{Astrophys. J.}
  {\bfseries 873} (Mar., 2019) 111},
  [\href{https://arxiv.org/abs/0805.2366}{{\ttfamily 0805.2366}}].

\bibitem{wfirst}
D.~{Spergel}, N.~{Gehrels}, C.~{Baltay}, D.~{Bennett}, J.~{Breckinridge},
  M.~{Donahue} et~al., \emph{{Wide-Field InfrarRed Survey
  Telescope-Astrophysics Focused Telescope Assets WFIRST-AFTA 2015 Report}},
  {\emph{arXiv e-prints} (Mar., 2015) arXiv:1503.03757},
  [\href{https://arxiv.org/abs/1503.03757}{{\ttfamily 1503.03757}}].

\bibitem{euclid}
R.~{Laureijs}, J.~{Amiaux}, S.~{Arduini}, J.~L. {Augu{\`e}res},
  J.~{Brinchmann}, R.~{Cole} et~al., \emph{{Euclid Definition Study Report}},
  {\emph{arXiv e-prints} (Oct., 2011) arXiv:1110.3193},
  [\href{https://arxiv.org/abs/1110.3193}{{\ttfamily 1110.3193}}].

\bibitem{takahashi/etal:2017}
R.~{Takahashi}, T.~{Hamana}, M.~{Shirasaki}, T.~{Namikawa}, T.~{Nishimichi},
  K.~{Osato} et~al., \emph{{Full-sky Gravitational Lensing Simulation for
  Large-area Galaxy Surveys and Cosmic Microwave Background Experiments}},
  \href{https://doi.org/10.3847/1538-4357/aa943d}{\emph{Astrophys. J.}
  {\bfseries 850} (Nov., 2017) 24},
  [\href{https://arxiv.org/abs/1706.01472}{{\ttfamily 1706.01472}}].

\bibitem{hamana/etal:2020}
T.~{Hamana}, M.~{Shirasaki}, S.~{Miyazaki}, C.~{Hikage}, M.~{Oguri}, S.~{More}
  et~al., \emph{{Cosmological constraints from cosmic shear two-point
  correlation functions with HSC survey first-year data}},
  \href{https://doi.org/10.1093/pasj/psz138}{\emph{Publ. Astron. Soc. Jap.}
  {\bfseries 72} (Feb., 2020) 16},
  [\href{https://arxiv.org/abs/1906.06041}{{\ttfamily 1906.06041}}].

\bibitem{chiang/etal:2013}
C.-T. Chiang et~al., \emph{{Galaxy redshift surveys with sparse sampling}},
  \href{https://doi.org/10.1088/1475-7516/2013/12/030}{\emph{JCAP} {\bfseries
  12} (2013) 030}, [\href{https://arxiv.org/abs/1306.4157}{{\ttfamily
  1306.4157}}].

\bibitem{alonso/etal:2014}
D.~Alonso, P.~G. Ferreira and M.~G. Santos, \emph{{Fast simulations for
  intensity mapping experiments}},
  \href{https://doi.org/10.1093/mnras/stu1666}{\emph{Mon. Not. Roy. Astron.
  Soc.} {\bfseries 444} (2014) 3183--3197},
  [\href{https://arxiv.org/abs/1405.1751}{{\ttfamily 1405.1751}}].

\bibitem{Xavier/etal:2016}
H.~S. Xavier, F.~B. Abdalla and B.~Joachimi, \emph{{Improving lognormal models
  for cosmological fields}}, \href{https://doi.org/10.1093/mnras/stw874,
  10.1093/mnras/459/4/3693}{\emph{Mon. Not. Roy. Astron. Soc.} {\bfseries 459}
  (2016) 3693--3710}, [\href{https://arxiv.org/abs/1602.08503}{{\ttfamily
  1602.08503}}].

\bibitem{Agrawal/etal:2017}
A.~Agrawal, R.~Makiya, C.-T. Chiang, D.~Jeong, S.~Saito and E.~Komatsu,
  \emph{{Generating Log-normal Mock Catalog of Galaxies in Redshift Space}},
  \href{https://doi.org/10.1088/1475-7516/2017/10/003}{\emph{JCAP} {\bfseries
  1710} (2017) 003}, [\href{https://arxiv.org/abs/1706.09195}{{\ttfamily
  1706.09195}}].

\bibitem{hand/etal:2018}
N.~Hand, Y.~Feng, F.~Beutler, Y.~Li, C.~Modi, U.~Seljak et~al.,
  \emph{{nbodykit: an open-source, massively parallel toolkit for large-scale
  structure}}, \href{https://doi.org/10.3847/1538-3881/aadae0}{\emph{Astron.
  J.} {\bfseries 156} (2018) 160},
  [\href{https://arxiv.org/abs/1712.05834}{{\ttfamily 1712.05834}}].

\bibitem{blot/etal:2019}
L.~Blot et~al., \emph{{Comparing approximate methods for mock catalogues and
  covariance matrices II: Power spectrum multipoles}},
  \href{https://doi.org/10.1093/mnras/stz507}{\emph{Mon. Not. Roy. Astron.
  Soc.} {\bfseries 485} (2019) 2806--2824},
  [\href{https://arxiv.org/abs/1806.09497}{{\ttfamily 1806.09497}}].

\bibitem{lippich/etal:2019}
M.~Lippich et~al., \emph{{Comparing approximate methods for mock catalogues and
  covariance matrices -- I. Correlation function}},
  \href{https://doi.org/10.1093/mnras/sty2757}{\emph{Mon. Not. Roy. Astron.
  Soc.} {\bfseries 482} (2019) 1786--1806},
  [\href{https://arxiv.org/abs/1806.09477}{{\ttfamily 1806.09477}}].

\bibitem{addison/etal:2019}
G.~Addison, C.~Bennett, D.~Jeong, E.~Komatsu and J.~Weiland, \emph{{The Impact
  of Line Misidentification on Cosmological Constraints from Euclid and other
  Spectroscopic Galaxy Surveys}},
  \href{https://doi.org/10.3847/1538-4357/ab22a0}{\emph{Astrophys. J.}
  {\bfseries 879} (2019) 15},
  [\href{https://arxiv.org/abs/1811.10668}{{\ttfamily 1811.10668}}].

\bibitem{sunayama/etal:2020}
T.~Sunayama, M.~Takada, M.~Reinecke, R.~Makiya, T.~Nishimichi, E.~Komatsu
  et~al., \emph{{Mitigating the impact of fiber assignment on clustering
  measurements from deep galaxy redshift surveys}},
  \href{https://doi.org/10.1088/1475-7516/2020/06/057}{\emph{JCAP} {\bfseries
  06} (2020) 057}, [\href{https://arxiv.org/abs/1912.06583}{{\ttfamily
  1912.06583}}].

\bibitem{coles/etal:1991}
P.~{Coles} and B.~{Jones}, \emph{{A lognormal model for the cosmological mass
  distribution.}}, \href{https://doi.org/10.1093/mnras/248.1.1}{\emph{Mon. Not.
  R. Astron. Soc.} {\bfseries 248} (Jan., 1991) 1--13}.

\bibitem{colombi:1994}
S.~{Colombi}, \emph{{A ``skewed'' lognormal approximation to the probability
  distribution function of the large-scale density field.}},
  \href{https://doi.org/10.1086/174834}{\emph{Astrophys. J.} {\bfseries 435}
  (Nov., 1994) 536--539},
  [\href{https://arxiv.org/abs/astro-ph/9402071}{{\ttfamily
  astro-ph/9402071}}].

\bibitem{kofman/etal:1994}
L.~{Kofman}, E.~{Bertschinger}, J.~M. {Gelb}, A.~{Nusser} and A.~{Dekel},
  \emph{{Evolution of One-Point Distributions from Gaussian Initial
  Fluctuations}}, \href{https://doi.org/10.1086/173541}{\emph{Astrophys. J.}
  {\bfseries 420} (Jan., 1994) 44},
  [\href{https://arxiv.org/abs/astro-ph/9311028}{{\ttfamily
  astro-ph/9311028}}].

\bibitem{bernardeau/kofman:1995}
F.~{Bernardeau} and L.~{Kofman}, \emph{{Properties of the Cosmological Density
  Distribution Function}},
  \href{https://doi.org/10.1086/175542}{\emph{Astrophys. J.} {\bfseries 443}
  (Apr., 1995) 479}, [\href{https://arxiv.org/abs/astro-ph/9403028}{{\ttfamily
  astro-ph/9403028}}].

\bibitem{uhlemann/etal:2016}
C.~{Uhlemann}, S.~{Codis}, C.~{Pichon}, F.~{Bernardeau} and P.~{Reimberg},
  \emph{{Back in the saddle: large-deviation statistics of the cosmic
  log-density field}}, \href{https://doi.org/10.1093/mnras/stw1074}{\emph{Mon.
  Not. R. Astron. Soc.} {\bfseries 460} (Aug., 2016) 1529--1541},
  [\href{https://arxiv.org/abs/1512.05793}{{\ttfamily 1512.05793}}].

\bibitem{shin/etal:2017}
J.~{Shin}, J.~{Kim}, C.~{Pichon}, D.~{Jeong} and C.~{Park}, \emph{{New Fitting
  Formula for Cosmic Nonlinear Density Distribution}},
  \href{https://doi.org/10.3847/1538-4357/aa74b9}{\emph{Astrophys. J.}
  {\bfseries 843} (July, 2017) 73},
  [\href{https://arxiv.org/abs/1705.06863}{{\ttfamily 1705.06863}}].

\bibitem{kayo/etal:2001}
I.~{Kayo}, A.~{Taruya} and Y.~{Suto}, \emph{{Probability Distribution Function
  of Cosmological Density Fluctuations from a Gaussian Initial Condition:
  Comparison of One-Point and Two-Point Lognormal Model Predictions with N-Body
  Simulations}}, \href{https://doi.org/10.1086/323227}{\emph{Astrophys. J.}
  {\bfseries 561} (Nov., 2001) 22--34},
  [\href{https://arxiv.org/abs/astro-ph/0105218}{{\ttfamily
  astro-ph/0105218}}].

\bibitem{monaco/2016}
P.~Monaco, \emph{{Approximate methods for the generation of dark matter halo
  catalogs in the age of precision cosmology}},
  \href{https://doi.org/10.3390/galaxies4040053}{\emph{Galaxies} {\bfseries 4}
  (2016) 53}, [\href{https://arxiv.org/abs/1605.07752}{{\ttfamily
  1605.07752}}].

\bibitem{colavincenzo/etal:2019}
M.~Colavincenzo et~al., \emph{{Comparing approximate methods for mock
  catalogues and covariance matrices \textendash{} III: bispectrum}},
  \href{https://doi.org/10.1093/mnras/sty2964}{\emph{Mon. Not. Roy. Astron.
  Soc.} {\bfseries 482} (2019) 4883--4905},
  [\href{https://arxiv.org/abs/1806.09499}{{\ttfamily 1806.09499}}].

\bibitem{Hamana:2001vz}
T.~Hamana and Y.~Mellier, \emph{{Numerical study of statistical properties of
  the lensing excursion angles}},
  \href{https://doi.org/10.1046/j.1365-8711.2001.04685.x}{\emph{Mon. Not. Roy.
  Astron. Soc.} {\bfseries 327} (2001) 169},
  [\href{https://arxiv.org/abs/astro-ph/0101333}{{\ttfamily
  astro-ph/0101333}}].

\bibitem{Bartelmann/Schneider:2001}
M.~Bartelmann and P.~Schneider, \emph{{Weak gravitational lensing}},
  \href{https://doi.org/10.1016/S0370-1573(00)00082-X}{\emph{Phys. Rept.}
  {\bfseries 340} (2001) 291--472},
  [\href{https://arxiv.org/abs/astro-ph/9912508}{{\ttfamily
  astro-ph/9912508}}].

\bibitem{Kilbinger:2015}
M.~{Kilbinger}, \emph{{Cosmology with cosmic shear observations: a review}},
  \href{https://doi.org/10.1088/0034-4885/78/8/086901}{\emph{Reports on
  Progress in Physics} {\bfseries 78} (Jul, 2015) 086901},
  [\href{https://arxiv.org/abs/1411.0115}{{\ttfamily 1411.0115}}].

\bibitem{loVerde/afshordi:2008}
M.~LoVerde and N.~Afshordi, \emph{{Extended Limber Approximation}},
  \href{https://doi.org/10.1103/PhysRevD.78.123506}{\emph{Phys. Rev. D}
  {\bfseries 78} (2008) 123506},
  [\href{https://arxiv.org/abs/0809.5112}{{\ttfamily 0809.5112}}].

\bibitem{FFTW05}
M.~Frigo and S.~G. Johnson, \emph{The design and implementation of {FFTW3}},
  {\emph{Proceedings of the IEEE} {\bfseries 93} (2005) 216--231}.

\bibitem{planck2015}
{\scshape Planck} collaboration, P.~A.~R. Ade et~al., \emph{{Planck 2015
  results. XIII. Cosmological parameters}},
  \href{https://doi.org/10.1051/0004-6361/201525830}{\emph{Astron. Astrophys.}
  {\bfseries 594} (2016) A13},
  [\href{https://arxiv.org/abs/1502.01589}{{\ttfamily 1502.01589}}].

\bibitem{class1}
J.~Lesgourgues, \emph{{The Cosmic Linear Anisotropy Solving System (CLASS) I:
  Overview}},  \href{https://arxiv.org/abs/1104.2932}{{\ttfamily 1104.2932}}.

\bibitem{class2}
D.~Blas, J.~Lesgourgues and T.~Tram, \emph{{The Cosmic Linear Anisotropy
  Solving System (CLASS) II: Approximation schemes}},
  \href{https://doi.org/10.1088/1475-7516/2011/07/034}{\emph{JCAP} {\bfseries
  1107} (2011) 034}, [\href{https://arxiv.org/abs/1104.2933}{{\ttfamily
  1104.2933}}].

\bibitem{shirasaki/etal:2015}
M.~{Shirasaki}, T.~{Hamana} and N.~{Yoshida}, \emph{{Probing cosmology with
  weak lensing selected clusters - I. Halo approach and all-sky simulations}},
  \href{https://doi.org/10.1093/mnras/stv1854}{\emph{Mon. Not. R. Astron. Soc.}
  {\bfseries 453} (Nov., 2015) 3043--3067},
  [\href{https://arxiv.org/abs/1504.05672}{{\ttfamily 1504.05672}}].

\bibitem{hamana/etal:2015}
T.~{Hamana}, J.~{Sakurai}, M.~{Koike} and L.~{Miller}, \emph{{Cosmological
  constraints from Subaru weak lensing cluster counts}},
  \href{https://doi.org/10.1093/pasj/psv034}{\emph{Publ. Astron. Soc. Jap.}
  {\bfseries 67} (June, 2015) 34},
  [\href{https://arxiv.org/abs/1503.01851}{{\ttfamily 1503.01851}}].

\bibitem{hu/jain:2004}
W.~Hu and B.~Jain, \emph{{Joint galaxy - lensing observables and the dark
  energy}}, \href{https://doi.org/10.1103/PhysRevD.70.043009}{\emph{Phys. Rev.
  D} {\bfseries 70} (2004) 043009},
  [\href{https://arxiv.org/abs/astro-ph/0312395}{{\ttfamily
  astro-ph/0312395}}].

\end{thebibliography}\endgroup
\end{document}